\documentclass[journal,10pt]{IEEEtran}

\ifCLASSINFOpdf
\else
\fi
\pdfoutput=1 

\usepackage{array}
\usepackage{algorithmic}
\usepackage{amsmath}
\usepackage{graphicx,float}
\usepackage[utf8]{inputenc}
\usepackage{enumerate}
\usepackage{color}
\usepackage{amsfonts}
\usepackage{cite}

\usepackage{algorithm}
\usepackage{algorithmic}
\begin{document}
\title{MIMO-OFDM-Based Massive Connectivity With Frequency 
Selectivity Compensation}
% was not built to address multiple paragraphs
%

%\author{Michael~Shell,~\IEEEmembership{Member,~IEEE,}
%        John~Doe,~\IEEEmembership{Fellow,~OSA,}
%        and~Jane~Doe,~\IEEEmembership{Life~Fellow,~IEEE}% <-this %}
% stops a space
\author{Wenjun Jiang,  
    Mingyang Yue, 
    Xiaojun Yuan,~\IEEEmembership{Senior Member,~IEEE}, and Yong 
    Zuo}
\maketitle

\begin{abstract}

In this paper, we study how to efficiently and reliably detect active devices and 
estimate their channels in a multiple-input multiple-output (MIMO) orthogonal 
frequency-division multiplexing (OFDM) based grant-free non-orthogonal 
multiple access (NOMA) system to enable massive machine-type communications 
(mMTC). First, by exploiting the correlation of the channel frequency responses in 
narrow-band mMTC, we propose a block-wise linear channel model. Specifically, the 
continuous OFDM subcarriers in the narrow-band are divided into several sub-blocks 
and a linear function with only two variables (mean and slope) is used to approximate 
the frequency-selective channel in each sub-block. This significantly reduces the 
number of variables to be determined in channel estimation and the sub-block 
number can be adjusted to reliably compensate the channel frequency-selectivity. 
Second, we formulate the joint active device detection and channel estimation in the 
block-wise linear system as a Bayesian inference problem. By exploiting the 
block-sparsity of the channel matrix, we develop an efficient turbo message passing 
(Turbo-MP) algorithm to resolve the Bayesian inference problem with near-linear 
complexity. We further incorporate machine learning approaches into Turbo-MP to 
learn unknown prior parameters. Numerical results demonstrate the superior 
performance of the proposed algorithm over state-of-the-art algorithms.

\end{abstract}

\begin{IEEEkeywords}
grant-free non-orthogonal multiple access, orthogonal frequency division multiplexing, 
turbo message passing.
\end{IEEEkeywords}

% \fi
\IEEEpeerreviewmaketitle

\section{Introduction}
\IEEEPARstart{m}{assive} machine-type communications (mMTC) has been envisioned as one of the three key application scenarios of fifth-generation (5G) wireless communications. To support massive connectivity of machine-type devices, mMTC typically has the feature of sporadic transmission with short packets, which is different from conventional human-type communications \cite{MTC}. This implies that only a small subgroup of devices are active in any time instance of mMTC. As such, in addition to channel estimation and signal detection, a fundamentally new challenge for the design of an mMTC receiver is to reliably and efficiently identify which subgroup of devices are actively engaged in packet transmission. 

Recently, a new random access protocol termed grant-free non-orthogonal 
multiple access (NOMA) has been evaluated and highlighted for mMTC 
\cite{TR38802}. In specific, in grant-free NOMA, the devices share the same 
time and frequency resources for signal transmission, and the signals can be 
transmitted without the scheduling grant from the base station (BS). The 
receiver at the BS is then required to perform active device detection (ADD), 
channel estimation (CE), and signal detection (SD), either separately or 
jointly. The earlier work  \cite{Bichai,Duyang,Jeong,Lei} assumed that full 
channel state information  (CSI) can be acquired at the BS and studied joint 
CE and SD. However, the assumption of full CSI availability is not practical 
since it will cause a huge overhead to estimate the CSI of all access devices. 
The follow-up work \cite{RAP} proposed to divide the process at the BS 
into two phases, namely, the joint ADD and CE phase the SD phase. Since 
the BS only needs to estimate the CSI of active devices, the pilot overhead is 
significantly reduced. In addition, the channel sparsity in the device domain 
enables the employment of compressed sensing (CS) algorithms \cite{CS} to 
solve the joint ADD and CE problem. For example, the authors in 
\cite{AMP_Liu} considered the multiple-input multiple-output (MIMO) 
transmission and leveraged a multiple measurement vector (MMV) 
CS technique termed vector approximate message passing 
(Vector AMP) \cite{AMP_Chen} to achieve asymptotically perfect ADD. In 
\cite{Turbo_Liu}, the authors further considered a mixed 
analog-to-digital converters (ADCs) architecture at the BS antennas 
and proposed a CS algorithm based on the turbo compressive sensing 
(Turbo-CS) \cite{Ma_1}. It is known from \cite{Ma_1} that Turbo-CS 
outperforms approximate message passing (AMP) \cite{AMP_1} both in 
convergence performance and computational complexity. Another line of 
research  considered the more challenging joint ADD, CE, and SD problem 
\cite{Qinyun,Tian}. As compared to the two-phase approach, these joint 
schemes can achieve significant performance improvement but at the expense 
of higher computational complexity.

Orthogonal frequency division multiplexing (OFDM) is a mature and enabling technology for 5G to provide high spectral efficiency. As such, the design of OFDM-based grant-free NOMA has attracted much research interest in recent years \cite{Yuanyuan,Zhaoji,Malong}. In \cite{Yuanyuan}, the authors exploited the block-sparsity of the channel responses on OFDM subcarriers to design a message-passing-based iterative algorithm. Besides, it has been demonstrated in \cite{Zhaoji} that the message-passing-based iterative algorithm can be unfolded into a deep neural network. By training the parameters of the neural network, the convergence and performance of the algorithm are improved. Furthermore, OFDM-based grant-free NOMA with massive MIMO has been considered in \cite{Malong}. By leveraging the sparsity both in the device domain and the virtual angular domain, the authors utilized the AMP algorithm to achieve the joint ADD and CE. One issue in \cite{Yuanyuan,Zhaoji,Malong} is that the frequency-domain channel estimation on every subcarrier requires a high pilot overhead, which is inefficient for the short data packets transmission in mMTC.

To reduce the pilot overhead, a common strategy is to transform the frequency-domain channel into the time-domain channel by inverse discrete Fourier transform (IDFT) \cite{STCS}. Due to the limited delay spread in the time-domain channel, the time-domain channel is sparse, thereby requiring fewer pilots for CE. Furthermore, by exploiting the sparsity of both the time-domain channel and the device activity pattern, some state-of-the-art CS algorithms such as Turbo-CS \cite{Ma_1,TARM} and Vector AMP \cite{AMP_Chen,AMP_Liu} can be applied to the considered systems with some straightforward modifications. However, there exists an energy leakage problem caused by the IDFT transform to obtain the time-domain channel. The energy leakage compromises the channel sparsity in the time domain. In addition, the power delay profile (PDP) is generally difficult for the BS to acquire, and thus cannot be exploited as prior information to improve the system performance.

Motivated by the bottleneck of the existing channel models when applied to the 
MIMO-OFDM-based grant-free NOMA system, we aim to construct a channel model 
to enable efficient massive random access. Due to the short packets in mMTC, the 
bandwidth for packet transmission is usually narrow, e.g., 1MHz for $10^5$ access 
devices \cite{LTE-M}. Then the variations of the channel frequency responses across 
the subcarriers are limited and slow. By leveraging this fact, we propose a block-wise 
linear channel model. Specifically, the continuous subcarriers are divided into several 
sub-blocks. In each sub-block, the frequency-selective channel is approximated by a 
linear function. We demonstrate that the number of variables in the block-wise linear 
channel model is typically much less than the number of non-zero delay taps in the 
time-domain channel. Moreover, the sub-block number can be modified to strike the 
trade-off between the model accuracy and the number of the channel variables to be 
estimated. 

Based on the block-wise linear system model, we aim to design a CS algorithm to solve the joint ADD and CE problem. We first introduce a probability model to characterize the block-sparsity of the channel matrix. Then the joint ADD and CE is formulated as a Bayesian inference problem. Inspired by the success of Turbo-CS \cite{Ma_1,TARM} in sparse signal recovery, we design a  message passing algorithm termed turbo message passing (Turbo-MP) to resolve the Bayesian inference problem. The message passing processes are derived based on the principle of Turbo-CS and the sum-product rule \cite{sum-product}. By designing a partial orthogonal pilot matrix with fast transform, Turbo-MP achieves near-linear complexity. 

Furthermore, we show that machine learning methods can be incorporated into 
Turbo-MP to learn the unknown prior parameters. We adopt the expectation 
maximization (EM) algorithm \cite{EM} to learn the prior parameters. We then show 
how to unfold Turbo-MP into a neural network (NN), where the prior parameters are 
seen as the learnable parameters of the neural network. Numerical results show that 
NN-based Turbo-MP has a faster convergence rate than EM-based Turbo-MP. More 
importantly, we show that Turbo-MP designed for the propose frequency-domain 
block-wise linear model significantly outperforms the state-of-the-art counterparts, 
especially those message passing algorithms designed for the time-domain sparse 
channel model \cite{AMP_Liu,Turbo_Liu}.

\subsection{Notation and Organization}
We use bold capital letters like $\mathbf X$ for matrices and bold lowercase 
letters $\mathbf x$ for vectors. $(\cdot)^T$ and $(\cdot)^H$ are used to 
denote the transpose and the conjugate transpose, respectively. We use ${\rm 
diag}(\mathbf x)$ for the diagonal matrix created from vector $\mathbf x$, ${\rm 
diag}(\mathbf x_1,...,\mathbf x_N)$ for the block diagonal matrix with the $n$-th 
block being vector $\mathbf x_n$, and ${\rm diag}(\mathbf X_1,...,\mathbf X_N)$ 
for the block diagonal matrix with the $n$-th block being matrix $\mathbf X_n$. We 
use ${\rm vec}(\mathbf X)$ for the vectorization of matrix $\mathbf X$ and 
$\otimes$ for the Kronecker product. $||\mathbf X ||_F$ and $||\mathbf x||_2$ are used 
to denote the Frobenius norm of matrix $\mathbf X$ and the $l_2$ norm of vector 
$\mathbf x$, respectively. Matrix $\mathbf I$ denotes the identity matrix with an 
appropriate size. For a random vector $\mathbf x$, we denote its probability density 
function (pdf) by $p(\mathbf x)$. $\delta(\cdot)$ denotes the Dirac delta function and 
$\delta[\cdot]$ denotes the Kronecker delta function. The pdf of a complex Gaussian 
random vector $\mathbf x \in \mathbb C^N$ with mean $\mathbf m$ and covariance 
$\mathbf C$ is denoted by $\mathcal{CN}(\mathbf x ; \mathbf m , \mathbf C) = {\rm 
exp}(-(\mathbf x - \mathbf m)^H \mathbf C^{-1} (\mathbf x - \mathbf m))/(\pi^N 
|\mathbf C|)$.

The remainder of this paper is organized as follows. In Section II, we introduce the existing system models. Furthermore, we propose the block-wise linear system model and demonstrate its superiority. In Section III, we formulate a Bayesian inference problem to address the ADD and CE problem. In Section IV, we propose the Turbo-MP algorithm, describe the pilot design, and analyze the algorithm complexity.  In Section V, we apply the EM algorithm to learn the prior parameters. Moreover, we show how Turbo-MP is unfolded into a neural network. In Section VI, we present the numerical results. In Section VII, we conclude this paper.

%\vspace{5pt}
%\begin{table}[h]
%	\small
%\begin{tabular}[c]{cc}
%	\hline
%	Notation & Definition  \\
%	\hline
%	$K$ & Number of access devices \\
%	$N$ & Number of OFDM subcarriers  \\
%	$M$ & Number of BS antennas \\
%	$L_k$ & Number of channel taps of the $k$-th device \\ 
%	$T$ & Number of OFDM symbols \\
%	$Q$ & Number of sub-blocks \\
%	\hline	
%	$\alpha_k$ & Indicator function of the $k$-th device \\
%	$\lambda$ & Device activity rate \\
%	\hline
%	$\mathbf Y$ & Observation matrix \\  
%	$\mathbf y_m$ & the $m$-th column of $\mathbf Y$ \\
%	$\mathbf H $ & Channel mean matrix \\
%	$\mathbf{h}_k$ & Rows $(k-1)Q+1,...,kQ,$ of $\mathbf H $ \\
%	$\mathbf b_{m}$ & the $m$-th column of $\mathbf H $  \\
%	$\mathbf C $ & Channel compensation matrix \\
%	$\mathbf{c}_k$ & Rows $(k-1)Q+1,...,kQ,$ of $\mathbf C $ \\
%	$\mathbf u_{m}$ & the $m$-th column of $\mathbf C $ \\
%	$\mathbf A$ & Pilot matrix \\
%	
%	\hline
%\end{tabular}
%\end{table}

\section{System Modeling} 

\subsection{MIMO-OFDM-Based Grant-free NOMA Model}

\begin{figure}[t]
	\centering
	\includegraphics[width=3.5in]{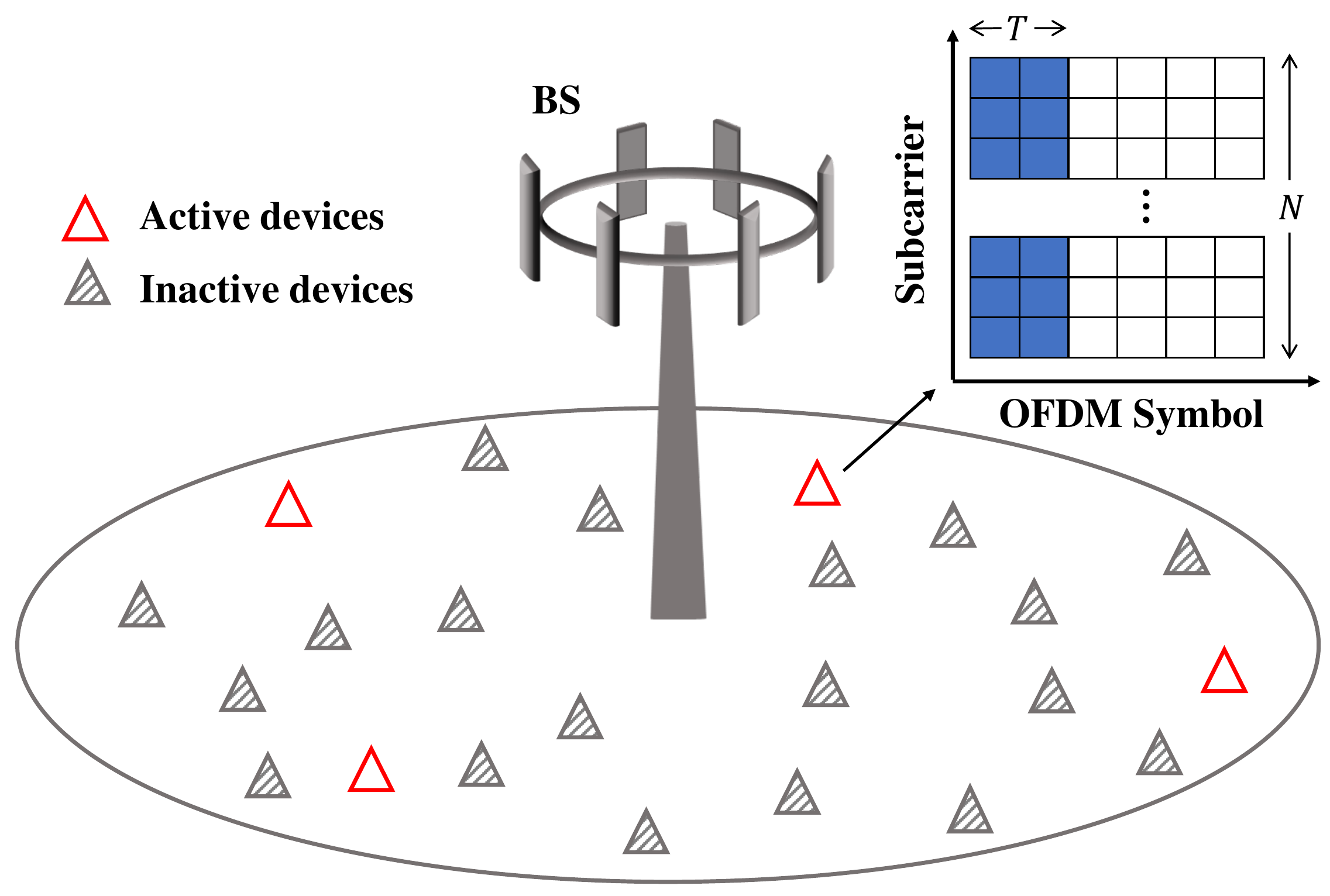}
	\caption{An illustration of MIMO-OFDM-based mMTC, where a small subgroup of devices are active in each time instance and share $N$ continuous 
	subcarriers for pilot transmission within $T$ OFDM symbols.}
	\label{MTC}
\end{figure}

Consider a MIMO-OFDM-based grant-free NOMA system as shown in Fig. 
\ref{MTC}, where a frequency band of $N$ adjacent OFDM subcarriers are 
allocated for mMTC. $N$ is usually much less than the total number of 
available subcarriers in the considered OFDM system. This frequency 
allocation strategy follows the idea of narrow-band internet-of-things 
\cite{Chen} and network slicing \cite{slice}. Then the allocated bandwidth is 
used to support $K$ single-antenna devices to randomly access an 
$M$-antenna base station (BS). In each time instance, only a small subset of 
devices are active. To characterize such sporadic transmission, the device 
activity is described by an indicator function 
$\alpha_k$ as
\begin{equation}
\alpha_{k}=\left
\{\begin{array}{l}
1, \text { device } k \text { is active } \\
0, \text { device } k \text { is inactive},
\end{array} \quad k=1,...,K \right.
\end{equation}
with $p(\alpha_k=1)=\lambda$ where $\lambda \ll 1$.

We adopt a multipath block-fading channel, i.e., the multipath channel response remain constant within the coherence time. Denote the channel frequency response on the $n$-th subcarrier from the $k$-th device at $m$-th BS antenna by
\begin{align}
g_{k,m,n} = & \sum\limits_{l = 1}^{L_k} \sqrt{\rho_{k,l}} 
\beta_{k,m,l} {e^{{ -j2  \pi \Delta f  \tau_{k,l} n}}}, \notag \\
& \quad k=1,...,K ; ~ m=1,...,M ; ~ n=1,...,N
\end{align} 
where $\Delta f$ is the OFDM subcarrier spacing; $L_k$ is the number of 
channel taps of the $k$-th device; $\rho_{k,l}$ and $\tau_{k,l} 
$\footnote{Strictly speaking, the differences of the tap delay at different BS 
antennas are absorbed in $\beta_{k,m,l}$} are respectively the $l$-th tap 
power and tap delay of the $k$-th device; $\beta_{k,m,l} \sim 
\mathcal{CN}(\beta_{k,m,l};0,1)$ is the normalized complex gain and 
assumed to be independent for any $k,m,l$ \cite{LiY}. Then the channel 
frequency response can be expressed in a matrix form as 
\begin{align} 
\mathbf G_k & =  \alpha_k
\left( \begin{array}{ccccc} 
g_{k,1,1} & \cdots & g_{k,m,1} & \cdots & g_{k,M,1} \\
\vdots & & \vdots & & \vdots \\
g_{k,1,N} & \cdots & g_{k,m,N} & \cdots & g_{k,M,N}
\end{array} \right)
\end{align} 

Let $a_{k,n}^{(t)}$ be the pilot symbol of the $k$-th device transmitted on th $n$-th subcarrier at the $t$-th OFDM symbol, and $T$ be the number of OFDM symbols for pilot transmission. Then we construct a block diagonal matrix $\mathbf \Lambda_k \in \mathbb C^{TN \times N}$ with the $n$-th diagonal block being $[a_{k,n}^{(1)},...,a_{k,n}^{(T)}]^T$, i.e.,

\begin{equation} \label{Lambda}
\mathbf \Lambda_k = {\rm 
diag}\big([a_{k,1}^{(1)},...,a_{k,1}^{(T)}]^T,...,[a_{k,N}^{(1)},..., 
a_{k,N}^{(T)}]^T \big)
\end{equation} 
Assume the cyclic-prefix (CP) length $L_{cp}> \tau_{k,l}, \forall k,l$. After removing the CP and applying the discrete Fourier transform (DFT), the system model in the frequency domain is described as
\begin{equation} \label{eq_mt}
\mathbf Y = \sum_{k=1}^K  \mathbf \Lambda_k \mathbf G_k + \mathbf N
\end{equation}
where $\mathbf Y \in \mathbb C^{TN \times M}$ is the observation matrix; $\mathbf N$ is an additive white Gaussian noise (AWGN) matrix with its elements independently drawn from $\mathcal{CN}(0,\sigma_N^2)$. In \cite{Yuanyuan,Zhaoji,Malong}, CS algorithms were proposed based on \eqref{eq_mt} to achieve the CE on every OFDM subcarrier.

\begin{figure}[t]
	\centering
	\includegraphics[width=3.5in]{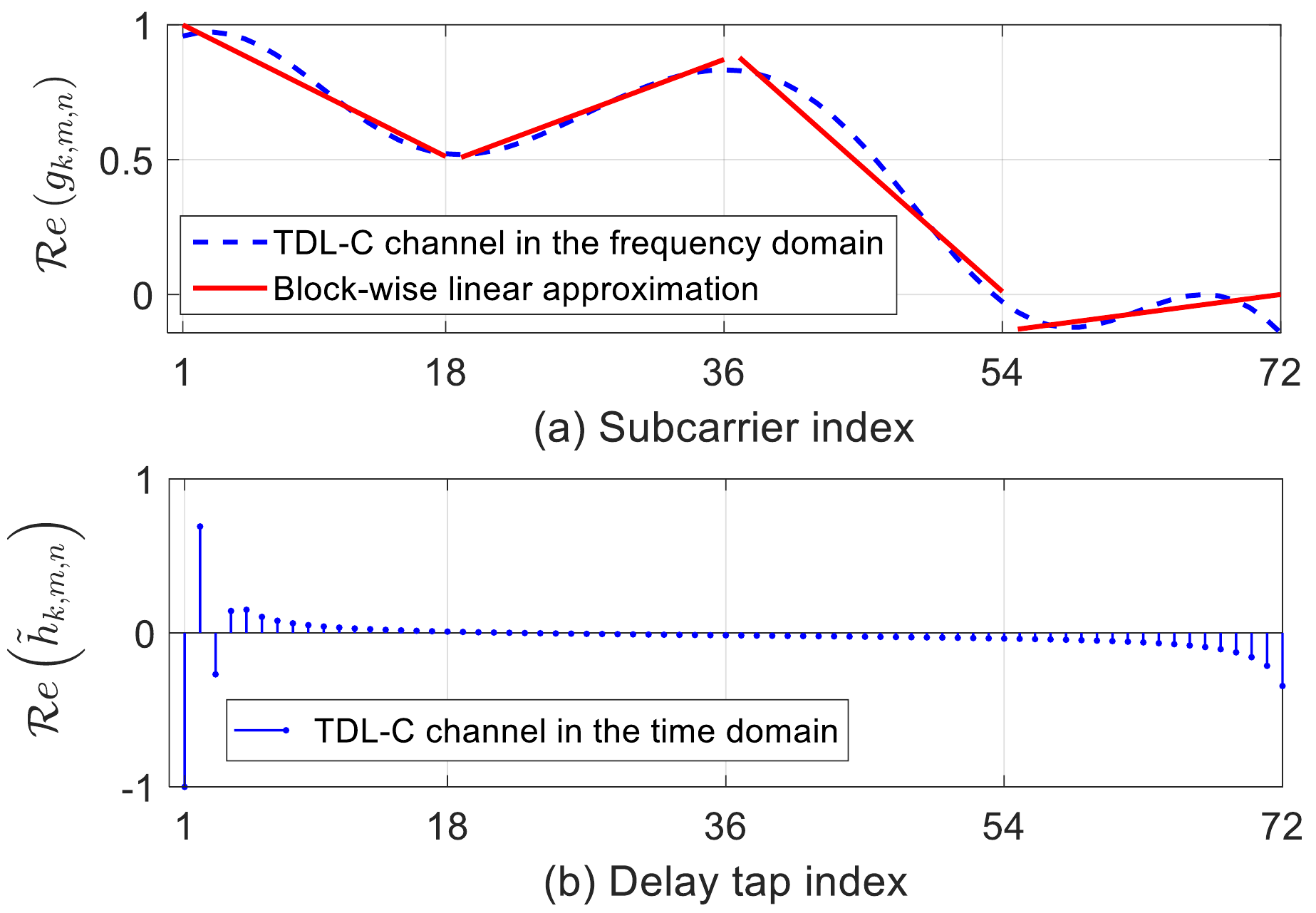}
	\caption{An example of channel model comparison. TDL-C multi-path channel in TR 38.901 R14 \cite{TR38901}. The number of OFDM subcarriers $N=72$. (a) frequency-domain channel response and its block-wise linear approximation with sub-block number $Q=4$. (b) corresponding time-domain channel response where the increase of the channel response at the tail is caused by the energy leakage of the IDFT transform.}
	\label{RB}
\end{figure}

Define the time-domain channel matrix of the $k$-th device as $\tilde{ \mathbf H}_{k} \in \mathbb C^{N \times M}$. Note that $\tilde{ \mathbf H}_{k}$ can be represented as the IDFT of $\mathbf G_k$, i.e,
\begin{align} \label{eq_Gk}
\tilde{\mathbf H}_{k} & = \mathbf F^H \mathbf G_k 
\end{align}
where $\mathbf F \in \mathbb C^{N \times N}$ is the DFT matrix with the $(n_1,n_2)$-th element being $1/\sqrt{N} \cdot {\rm exp}(-j2 \pi n_1 n_2/N)$. Similarly, $\mathbf G_k$ is represented as $\mathbf G_k = \mathbf F \tilde{ \mathbf H}_{k}$. Substituting $\mathbf G_k = \mathbf F \tilde{ \mathbf H}_{k}$ into \eqref{eq_mt}, we obtain 
\begin{align} \label{eq_mt2}
\mathbf Y & = \sum_{k=1}^K  \mathbf \Lambda_k \mathbf F \tilde{ \mathbf H}_{k} + 
\mathbf N \notag \\
& = [\mathbf \Lambda_1 \mathbf F,...,\mathbf \Lambda_K \mathbf F] 
\tilde{\mathbf H} + \mathbf N
\end{align}
where $\tilde{ \mathbf H} = [\tilde{\mathbf H}_{1}^T,...,\tilde{\mathbf H}_{K}^T]^T $ is the channel matrix in the time domain. It is known that the channel delay spread is usually much smaller than $N$, i.e., some rows of $\tilde{\mathbf H}_{k}$ are zeros. Besides, due to the sporadic transmission of the devices, $\tilde{ \mathbf H}_{k}$ is an all-zero matrix if device $k$ is inactive. In this case, CS algorithms such as Vector AMP \cite{AMP_Chen,AMP_Liu} and Turbo-CS \cite{Ma_1,TARM} can be used to recover $\tilde{\mathbf H}$ from observation $\mathbf Y$ by exploiting the sparsity of $\tilde{\mathbf H}$. However, path delay is generally not an integer multiple of the sampling interval, resulting in the energy leakage problem of the IDFT transform \eqref{eq_Gk} which severely compromises the sparsity of $\tilde{\mathbf H}$. An illustration of the energy leakage problem is given in Fig. \ref{RB}(b).

%The channel response in the time-domain are shown in Fig. \ref{RB}(b) as 
%an example. It is clear that the number of non-zero elements of the channel 
%response is less than the subcarirers number. However, the energy 
%difference 
%is large between the taps in both ends and those in the middle, making the 
%taps in the middle difficult to estimate. In the next subsection, we propose 
%a 
%model in the frequency domain to achieve the effective estimation.

\subsection{Proposed Block-Wise Linear System Model}

For narrow-band mMTC  \cite{MTC}, the variations of the channel 
frequency response across the subcarriers are typically slow and limited, 
which inspires us to develop an alternative channel model to efficiently 
leverage the correlation in the frequency domain. In specific, we propose a 
block-wise linear channel model. We divide the $N$ continuous subcarriers 
into $Q$ sub-blocks. In each sub-block $q$, a linear function is used as the 
approximation of the channel frequency response:
\begin{align} \label{eq_linear_Apr}
 g_{k,n_q,m} = h_{k,q,m} & + (n_q-l_q) c_{k,q,m} + \Delta_{k,n_q,m} \notag \\
  & n_q =(q-1) N/Q+1,...,q N/Q
\end{align} 
where $h_{k,q,m}$ and $c_{k,q,m}$ represent the mean and slope of the 
linear function in the $q$-th sub-block, respectively; $\Delta_{k,n_q,m}$ is the error term due to model mismatch; and $l_q = (q-\frac{1}{2})N/Q$ is the midpoint of $n_q$. Intuitively, $h_{k,q,m}$ can be seen as the mean-value of the channel 
response in the $q$-th sub-block, and $c_{k,q,m}$ is used  to characterize 
the change of the channel response for the compensation of the 
frequency-selectivity. For the $k$-th device,  we define the matrix $\mathbf 
H_k \in \mathbb C^{Q\times M}$ and $\mathbf C_k \in \mathbb 
C^{Q\times M}$ as
\begin{align} 
\label{eq_Hk}
\mathbf H_k & =  \alpha_k
\left( \begin{array}{ccccc} 
h_{k,1,1} & \cdots & h_{k,1,m} & \cdots & h_{k,1,M} \\
\vdots & & \vdots & & \vdots \\
h_{k,Q,1} & \cdots & h_{k,Q,m} & \cdots & h_{k,Q,M}
\end{array} \right)  \\
\label{eq_Ck}
\mathbf C_k & = \alpha_k 
\left( \begin{array}{ccccc} 
c_{k,1,1} & \cdots & c_{k,1,m} & \cdots & c_{k,1,M} \\
\vdots & & \vdots & & \vdots \\
c_{k,Q,1} & \cdots & c_{k,Q,m} & \cdots & c_{k,Q,M}
\end{array} \right)_. 
\end{align} 
The reason for introducing $\alpha_k$ in (\ref{eq_Hk})-(\ref{eq_Ck}) is 
that the channel estimation and device activity detection can be jointly 
achieved by recovering $\mathbf H_k$ and $\mathbf C_k$. 

Define  $\mathbf E_1 = {\rm diag}(\mathbf 1_{N/Q},...,\mathbf 1_{N/Q}) 
\in \mathbb R^{N\times Q}$ with $\mathbf 1_{N/Q}$ being an all-one vector of length $N/Q$ and $\mathbf E_2 = {\rm diag}(\mathbf d,...,\mathbf d) \in \mathbb R^{N\times Q} $ with $\mathbf d = [-\frac{N}{2Q}+1,...,\frac{N}{2Q}]^T$. Then the block-wise linear approximation of the channel frequency response $\mathbf G_k$ is given by
\begin{equation} \label{eq_bl}
	\mathbf G_k = \mathbf E_{1} \mathbf H_k + \mathbf E_{2} \mathbf C_k + \mathbf \Delta_k
\end{equation}
where $\mathbf \Delta_k  \in \mathbb C^{N \times M}$ is the error matrix 
from the $k$-th device with the $(n,m)$-th element $\Delta_{k,n,m}$ in 
\eqref{eq_linear_Apr}. Substituting \eqref{eq_bl} into \eqref{eq_mt} with 
some manipulations, we obtain the block-wise linear system model as
\begin{align}
\label{eq_model}
\mathbf Y = \mathbf A \mathbf H + \mathbf B \mathbf C + 
          \mathbf W
\end{align}
where $\mathbf A = [\mathbf \Lambda_1 \mathbf E_1,...,\mathbf 
\Lambda_K \mathbf E_1] \in \mathbb C^{TN \times QK}$ and $\mathbf B 
= [\mathbf \Lambda_1 \mathbf E_2,...,\mathbf \Lambda_K \mathbf E_2] \in 
\mathbb C^{TN \times QK}$ are the pilot matrices; $\mathbf H = [\mathbf 
H_1^T,...,\mathbf H_K^T]^T \in \mathbb C^{QK \times M}$ is the 
\textit{channel mean matrix}; $\mathbf C = [\mathbf C_1^T,...,\mathbf 
C_K^T]^T \in \mathbb C^{QK \times M}$ is the \textit{channel 
compensation matrix}; $\mathbf W$ is the summation of the AWGN and the error terms from model mismatch as
\begin{equation}
\mathbf W = \sum_{k=1}^K \mathbf \Lambda_k \mathbf \Delta_k + \mathbf 
N.
\end{equation}
Following the central limit theorem, for a large devices number $K$, $\mathbf W$ can be modeled as an AWGN matrix with mean zero and variance $\sigma_w^2$.  We note that with $Q=N$ and $\mathbf C = \mathbf 0$, system model (\ref{eq_model}) reduces to the model \eqref{eq_mt} in \cite{Yuanyuan,Zhaoji,Malong} where the channel response on every subcarrier needs to be estimated exactly. Furthermore, we adjust the sub-block number $Q$ to strike a balance between the number of channel variables to be estimated and the model accuracy. An example is shown in Fig. \ref{RB} where the channel response in the frequency domain and its block-wise linear approximation is given in Fig. \ref{RB}(a). It is seen that sub-block number $Q=4$ is sufficient to ensure that the block-wise linear model approximates the frequency-domain channel response very well. This implies that, with model \eqref{eq_model}, only $2Q$ variables need to be estimated for each device at each BS antenna. Compared to the channel response in the time domain as shown in Fig. \ref{RB}(b), it is clear that the number of unknown variables $2Q$ is much less than the number of corresponding non-zero taps. In the remainder of the paper, we focus on the estimation algorithm design based on model (\ref{eq_model}). In the simulation section, we will show that our algorithm designed based on (\ref{eq_model}) significantly outperforms the existing approaches based on \eqref{eq_mt} and (\ref{eq_mt2}).

%Compared with the time-domain model (\ref{eq_mt2}) which exploits the limited delay spread to reduce the number of unknown channel variables, the proposed model (\ref{eq_model}) exploits the channel correlation in the frequency domain to reduce the number of channel variables to be estimated. We take an example to illustrate the superiority of the frequency-domain model. As shown in Fig. \ref{RB}(a), an effective linear approximation is obtained by setting sub-block number $Q=4$, which means only $8$ variables need to be estimated for the $k$-th device at the $m$-th BS antenna. However, as shown in Fig. \ref{RB}(b), the number of non-zeros taps is clearly larger than $8$. Besides, it is known that the channel taps in the middle have much lower energies than those in both ends. But in the frequency domain, although there is frequency-selective fading, the channel on each subcarrier has the same signal-to-noise ratio (SNR) in statistics. Especially when the BS is equipped with multiple antennas, the frequency-selective fading is significantly reduced. This suggests that the proposed frequency-domain model with block-wise liner approximation is more proper and useful for the OFDM-based mMTC. In the remainder of the paper, we focus on the detection algorithm design based on the model (\ref{eq_model}). In the simulation section, we will also show that our algorithm designed based on (\ref{eq_model}) can significantly outperform the existing approaches based on (\ref{eq_mt2}).  

\section{Problem Statement} 
With model \eqref{eq_model}, our goal is to recover the channel mean 
matrix $\mathbf H$ and channel compensation matrix $\mathbf C$ based on 
the noisy observation $\mathbf Y$. This task can be constructed as a 
Bayesian inference problem. In the following, we first introduce the 
probability model of $\mathbf H$ and $\mathbf C$, and then describe the 
statistical inference problem.

Due to the sporadic transmission of the devices, matrices $\mathbf H$ and $\mathbf C$ have a structured sparsity referred to as block-sparsity. In specific, if the $k$-th device is inactive, we have $\mathbf H_k =\mathbf 0 $ and $\mathbf C_k = \mathbf 0$. With some abuse of notation, we utilize a conditional Bernoulli-Gaussian (BG) distribution \cite{STCS} to characterize the block-sparsity as 
\begin{align}
p(\mathbf{H}_{k} | \alpha_k) & \sim 
\delta[\alpha_k]\delta(\mathbf{H}_k)+\delta[1-\alpha_k] {\cal 
CN}(\mathbf{H}_k; \mathbf 0; \vartheta_{ \mathbf H} \mathbf I)
\notag \\ 
p(\mathbf{C}_{k} | \alpha_k) & \sim 
\delta[\alpha_k]\delta(\mathbf{C}_k)+\delta[1-\alpha_k] {\cal 
CN}(\mathbf{C}_k; 
\mathbf 0; \vartheta_{ \mathbf C} \mathbf I)  
\end{align}
where $\delta(\cdot)$ is the Dirac delta function and $\delta[\cdot]$ is the 
Kronecker delta function. With indicator function 
$\alpha_k=0$, $\mathbf{H}_k$ and $\mathbf{C}_k$ are both zeros. With 
$\alpha_k=1$, the elements of $\mathbf{H}_k$ and $\mathbf{C}_k$ are 
independent and identically distributed (i.i.d.) Gaussian with variances $\vartheta_{\mathbf H}$ and $\vartheta_{ \mathbf C}$, respectively. We further assume that each device accesses the BS in an i.i.d. manner. Then the indicator function $\alpha_k$ is drawn from the Bernoulli distribution as 
\begin{equation}  
p(\alpha_k) = (1-\lambda)\delta[\alpha_k] + \lambda \delta[1-\alpha_k]
\end{equation}
where $\lambda$ is the device activity rate. 

Consider an estimator to minimize the mean-square error (MSE) of $\mathbf H$ and $\mathbf C$. It is known that the estimator which minimizes the MSE is the posterior expectation with respect to the posterior distribution \cite{Kay}. Define $\mathbf h_m \in \mathbb C^{QK}$ and $\mathbf c_m \in \mathbb C^{QK}$ as the $m$-th column of $\mathbf H$ and $\mathbf C$, respectively. Define $\mathbf y_m \in \mathbb C^{TN}$ as the $m$-th column of $\mathbf Y$. Then the posterior distribution $p(\mathbf H , \mathbf C, \boldsymbol{\alpha} | \mathbf Y)$ is described as
\begin{align} \label{pd}
 & p(\mathbf H , \mathbf C,  \boldsymbol{\alpha} | \mathbf Y)  \propto 
 p(\mathbf Y | \mathbf H ,  \mathbf C )	p(\mathbf H , 
 \mathbf C, \boldsymbol{\alpha}) \notag \\
& \quad \propto \prod_{m} p(\mathbf y_m | \mathbf h_m , \mathbf c_m) \prod_k 
p(\mathbf H_k | \alpha_k) p(\mathbf C_k | \alpha_k) p(\alpha_k)
\end{align}
where $ \prod_{m} 
p(\mathbf y_m | \mathbf h_m , \mathbf c_m)$ and $\prod_k 
p(\mathbf H_k | \alpha_k) p(\mathbf C_k | \alpha_k) p(\alpha_k)$ are the 
likelihood and the prior, respectively; vector 
$\boldsymbol{\alpha}=[\alpha_1,...,\alpha_k]^T$. In mMTC with a large 
device number $K$, it is computationally intractable to obtain the minimum 
mean-square error (MMSE) estimator. In the following section, we propose a 
low-complexity algorithm termed turbo message passing (Turbo-MP) to 
obtain an approximate MMSE solution.

\section{Turbo Message Passing}

\subsection{Algorithm Framework}
\begin{figure*}[t]
\centering
\includegraphics[width=6in]{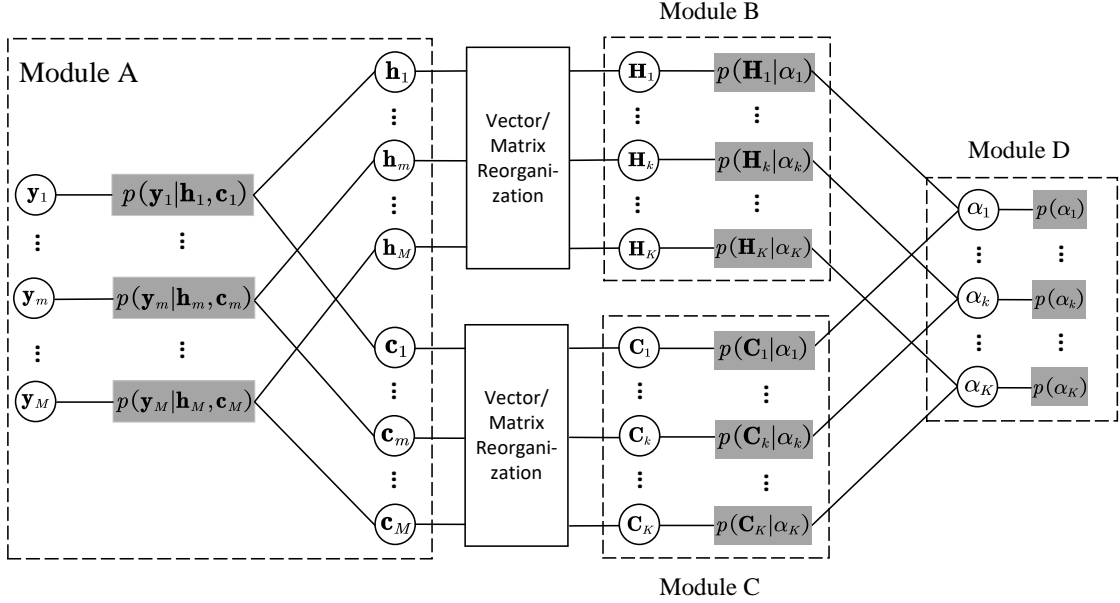}
\caption{The block diagram of the proposed turbo message passing (Turbo-MP) algorithm.}
\label{Turbo-CS}
\end{figure*} 

% More precisely with insight
% distributed
The factor graph representation of $ p(\mathbf H, \mathbf C, 
\boldsymbol{\alpha} | \mathbf Y)$ is shown in Fig. \ref{Turbo-CS}, based on which Turbo-MP is established. In the factor graph, the likelihood and prior as in \eqref{pd} are treated as the factor nodes (grey rectangles), while the random variables are treated as the variable nodes (blank circles). In specific, Turbo-MP consists of four modules referred to as Module A, B, C, and D, respectively. Module A is to obtain the estimates of the channel mean matrix $\mathbf H$ and the channel compensation matrix $\mathbf C$ by exploiting the likelihood $ \prod_{m} p(\mathbf y_m | \mathbf h_m , \mathbf c_m)$. Modules B and C are to obtain the estimates of  $\mathbf H$ and $\mathbf C$ by exploiting the priors $\prod_k p(\mathbf H_k | \alpha_k)$ and $p(\mathbf C_k | \alpha_k)$, respectively. Module D is to obtain the estimate of $\alpha_k$ by exploiting the prior $p(\alpha_k)$. The estimates are passed between modules in each Turbo-MP iteration like turbo decoding \cite{turbo}. For example, the output of Module B is used as the input of Module A, and vice versa. The four modules are executed iteratively until convergence. The derivation of Turbo-MP algorithm follows the sum-product rule \cite{sum-product} and the principle of Turbo-CS \cite{Ma_1,TARM}.

\subsection{Module A: Linear Estimation of $\mathbf h_m$ and $\mathbf c_m$}
In Module A, $\mathbf h_m$ and $\mathbf c_m$ at antenna $m=1,..,M$ are 
estimated separately by exploiting the  likelihood $ \prod_{m} p(\mathbf 
y_m | \mathbf h_m , \mathbf c_m)$ and the messages from Modules B and C.
In specific, denote the message from variable node $\mathbf h_m$ to  
factor node $p(\mathbf y_m|\mathbf h_m,\mathbf c_m)$ by 
$\mathcal{M}_{\mathbf h_m \to \mathbf y_m}(\mathbf h_m)$ and the 
message from variable node $\mathbf c_m$ to factor node $p(\mathbf 
y_m|\mathbf h_m,\mathbf c_m)$ by $\mathcal{M}_{\mathbf c_m \to 
\mathbf y_m}(\mathbf c_m)$. Following the principle of Turbo-CS
\cite{Ma_1,TARM}, we assume that the above two messages are Gaussian, 
i.e.,  $\mathcal{M}_{\mathbf h_m \to \mathbf y_m}(\mathbf h_m) \sim 
\mathcal{CN}(\mathbf h_m ; \mathbf h_m^{A,pri},v_{\mathbf 
h_m}^{A,pri} \mathbf I)$ and $\mathcal{M}_{\mathbf c_m \to \mathbf 
y_m}(\mathbf c_m) \sim \mathcal{CN}(\mathbf c_m ; \mathbf 
c_m^{A,pri},v_{\mathbf c_m}^{A,pri} \mathbf I)$. Note that the algorithm starts with $\mathbf h_m^{A,pri} = \mathbf 0 $, $\mathbf c_m^{A,pri} = \mathbf 0$, $v_{\mathbf h_m}^{A,pri} = \lambda \vartheta_{\mathbf H}$, and $v_{\mathbf c_m}^{A,pri} = \lambda \vartheta_{\mathbf C}$. From the sum-product rule, the belief of $h_{k,q,m}$ at factor node 
$p(\mathbf y_m|\mathbf h_m,\mathbf c_m)$ is
\begin{align} \label{eq_m_h_kmq}
\mathcal{M}_{\mathbf y_m}(h_{k,q,m}) = \int_{/ h_{k,q,m}} p(\mathbf y_m|\mathbf 
h_m,&\mathbf c_m)  \mathcal{M}_{\mathbf h_m \to \mathbf y_m}(\mathbf h_m) 
\notag \\ 
\times & \mathcal{M}_{\mathbf c_m \to \mathbf y_m}(\mathbf c_m).
\end{align}
In the above, $\mathbf y_m$ is the subscript is the shorthand for factor node 
$p(\mathbf y_m|\mathbf h_m,\mathbf c_m)$; $/ h_{k,q,m}$ denotes the set 
includes all elements in $\mathbf h_m$ except $ h_{k,q,m}$. Then we 
define the belief of $\mathbf h_{m}$ at factor node $p(\mathbf y_m|\mathbf 
h_m,\mathbf c_m)$ as
\begin{equation} \label{eq_m_h_m}
\mathcal{M}_{\mathbf y_m}(\mathbf h_{m}) = \prod_{k,q} \mathcal{M}_{\mathbf 
y_m}(h_{k,q,m}).
\end{equation}
Combing \eqref{eq_m_h_kmq} and \eqref{eq_m_h_m}, the message 
$\mathcal{M}_{\mathbf y_m}( \mathbf h_{m})$ is also Gaussian with the 
mean $\mathbf h_{m}^{A,post}$ and variance $v_{\mathbf 
h_m}^{A,pri}$. From \cite[Eq. 11]{TARM}, the mean $\mathbf 
h_{m}^{A,post}$ is expressed as
\begin{align} \label{hm}
\mathbf h_{m}^{A,post} = \  \mathbf h_{m}^{A,pri} & + \  v_{\mathbf 
h_m}^{A,pri} \mathbf A^H \mathbf {\boldsymbol \Sigma}_m^{-1}  \notag \\ 
& ~~~~ \times  \left(\mathbf y_m - \mathbf{A} \mathbf h_{m}^{A,pri} -\mathbf B 
\mathbf c_{m}^{A,pri} \right)
\end{align} 
where $\mathbf {\boldsymbol \Sigma}_m$ is the covariance matrix given by
\begin{equation}
\mathbf {\boldsymbol \Sigma}_m  = v_{\mathbf h_m}^{A,pri}\mathbf{AA}^H +  
v_{\mathbf c_m}^{A,pri} \mathbf{BB}^H + \sigma_w^2 \mathbf I.	
\end{equation}
To reduce the computational complexity of channel inverse ${\boldsymbol \Sigma}_m^{-1}$, we require that $\mathbf A$ is partial orthogonal \cite{Ma_1}, i.e., $\mathbf A \mathbf A^H = KP \mathbf I$, where $P$ is the average power of the pilot symbol. (The design of $\mathbf A$ to guarantee partial orthogonality is presented in \ref{pilot}.) In addition, it is easy to verify $\mathbf B = \mathbf D \mathbf A$ where the diagonal matrix $\mathbf D = {\rm diag}\big( [\mathbf d^T,...,\mathbf d^T]^T \otimes \mathbf (\mathbf 1_T)^T \big) \in \mathbb{R}^{TN \times TN}$ with $\mathbf 1_T$ being an all-one vector of length $T$. Then the expression of $\mathbf {\boldsymbol \Sigma}_m$ is simplified to
\begin{equation} \label{eq_C}
{\boldsymbol \Sigma}_m  = KP v_{\mathbf h_m}^{A,pri}\mathbf{I}+ KP 
v_{\mathbf c_m}^{A,pri} \mathbf{D} \mathbf{D}^H + \sigma_w^2 \mathbf I.
\end{equation}
Then the variance $v_{\mathbf h_m}^{A,post}$ is given by
\begin{equation} \label{hv}
v_{\mathbf h_m}^{A,post} = v_{\mathbf h_m}^{A,pri} - 
\sum_{i=1}^{TN} \frac{ P(v_{\mathbf h_m}^{A,pri})^2 }{ 
{\Sigma}_{m,i,i}  }
\end{equation} 
where ${\Sigma}_{m,i,i} $ is the $(i,i)$-th element of ${\boldsymbol 
\Sigma}_m $. We note that \eqref{hm} and \eqref{hv} also corresponds 
to the linear minimum mean-square error (LMMSE) estimator \cite[Chap. 
11]{Kay} given the observation $\mathbf y_m$, the mean $\mathbf 
h_m^{A,pri},\mathbf c_m^{A,pri}$and the variance $ v_{\mathbf 
h_m}^{A,pri},  v_{\mathbf c_m}^{A.pri} $. 

From the sum-produce rule, the extrinsic message from factor node 
$p(\mathbf y_m|\mathbf h_m \mathbf c_m)$ to variable node $\mathbf 
h_m$ is calculated as 
\begin{equation}
\mathcal{M}_{\mathbf y_m \to \mathbf h_m}(\mathbf h_m) \propto
\frac{\mathcal{M}_{\mathbf y_m}(\mathbf h_{m})}{\mathcal{M}_{\mathbf h_m \to 
\mathbf y_m}(\mathbf h_m)}.
\end{equation}
Given the Gaussian messages $\mathcal{M}(\mathbf h_{m})$ and 
$\mathcal{M}_{\mathbf h_m \to \mathbf y_m}(\mathbf h_{m})$, the extrinsic message is also Gaussian as 
\begin{equation}
\mathcal{M}_{\mathbf y_m \to \mathbf h_m}(\mathbf h_m) \sim 
\mathcal{CN}(\mathbf h_m ; \mathbf h_m^{A,ext}, v_{\mathbf 
h_m}^{A,ext} \mathbf I)
\end{equation}
where the variance $v_{\mathbf h_m}^{A,ext}$ and the mean $\mathbf 
h_m^{A,ext}$ are respectively given by
\begin{align}
&{v}_{\mathbf h_m}^{A,ext}=\left(\frac{1}{{v}_{\mathbf 
h_m}^{A,post}}-\frac{1}{{v}_{\mathbf h_m}^{A,pri}}\right)^{-1}  \notag 
\\
&\mathbf{h}_{m}^{A,ext}={v}_{\mathbf 
h_m}^{A,ext}\left(\frac{\mathbf{h}_{m}^{A,post}}{{v}_{\mathbf 
h_m}^{A,post}}-\frac{\mathbf{h}_{m}^{A,pri}}{{v}_{\mathbf h_m}^{A,pri}}\right). 
\end{align}

The calculation to obtain the belief of $\mathbf c_m$ is similar. We have 
the Gaussian belief $\mathcal{M}(\mathbf c_{m})$ with its mean and 
variance respectively given by 
\begin{align}
\mathbf c_{m}^{A,post} = \mathbf c_{m}^{A,pri} & + \  v_{\mathbf c_m}^{A,pri} 
\mathbf B^H \mathbf {\boldsymbol \Sigma}_m^{-1} \notag  \\ 
& \times \left(\mathbf y_m - \mathbf{A} \mathbf h_{m}^{A,pri} - \mathbf B  \mathbf c_{m}^{A,pri} \right) \notag \\
v_{\mathbf c_m}^{A,post} = v_{\mathbf c_m}^{A,pri}&  -  
\sum_{i=1}^{TN}  \frac{P D_{i,i}^2 (v_{\mathbf c_m}^{A,pri})^2  } { 
{\Sigma}_{m,i,i} }
\end{align}
where $D_{i,i}$ is the $(i,i)$-th element of $\mathbf D$. Then we obtain 
the extrinsic message $\mathcal{M}_{\mathbf y_m \to \mathbf 
c_m}(\mathbf c_m) \sim \mathcal{CN}(\mathbf c_m ; \mathbf 
c_m^{A,ext}, v_{\mathbf c_m}^{A,ext} \mathbf I)$ with its mean and 
variance as follows:
\begin{align}
&{v}_{\mathbf c_m}^{A,ext}=\left(\frac{1}{{v}_{\mathbf 
c_m}^{A,post}}-\frac{1}{{v}_{\mathbf c_m}^{A,pri}}\right)^{-1} \notag 
\\
&\mathbf{c}_{m}^{A,ext}={v}_{\mathbf c_m}^{A,ext} \left(\frac{\mathbf{c}_{m}^{A,post}}{{v}_{\mathbf c_m}^{A,post}}-\frac{\mathbf{c}_{m}^{A,pri}}{{v}_{\mathbf c_m}^{A,pri}}\right).
\end{align}

\subsection{Module B: Denoiser of $\mathbf H_k$}
In Module B, each $\mathbf H_k, \forall k$ is estimated individually by 
exploiting the prior $ p(\mathbf H_k|\alpha_k)$ and the messages from 
Modules A and D. In specific, given the message $\mathcal{M}_{\mathbf 
y_m \to \mathbf h_m}(\mathbf h_m) \sim \mathcal{CN}(\mathbf h_m ; 
\mathbf h_m^{A,ext}, v_{\mathbf h_m}^{A,ext} \mathbf I)$ from module 
A, we have 
\begin{equation}
	\mathcal{M}_{\mathbf y_m \to h_{k,q,m}}(h_{k,q,m}) \sim \mathcal{CN}( h_{k,q,m}; \mathbf h_{k,q,m}^{ext}, v_{\mathbf h_m}^{A,ext}).
\end{equation}
For description convenience, the factor node $p(\mathbf H_k | \alpha_k)$ is 
replaced by $f_k^B$. Then the message from variable node $\mathbf H_k$ 
to factor node $p(\mathbf H_k | \alpha_k)$ is expressed as
\begin{align}
\mathcal{M}_{\mathbf H_k \to f_k^B}(\mathbf H_k) & = 
\prod_{q,m} \mathcal{M}_{\mathbf y_m \to h_{k,q,m}}(h_{k,q,m}) \notag \\
& = {\cal{CN}} (\mathbf H_k ; \mathbf H_k^{B,pri} ,  \mathbf V^B )  
\end{align}
where $\mathbf H_k^{B,pri} \in \mathbb C^{Q\times M}$ with the 
$(q,m)$-th element $h_{k,q,m}^{B,pri} = h_{k,q,m}^{A,ext}$ and 
$\mathbf V^B = {\rm diag}([v_{\mathbf h_1}^{B,pri},...,v_{\mathbf 
h_M}^{B,pri}]^T \otimes \mathbf 1_Q^T )$ with $\mathbf 1_Q$ being an 
all-one vector of length $Q$. 

Combing the Bernoulli Gaussian prior $ p(\mathbf H_k|\alpha_k)$, 
Gaussian message $\mathcal{M}_{\mathbf H_k \to f_k^B}(\mathbf H_k)$ 
and Bernoulli message $\mathcal{M}_{\alpha_k \to f^B_{k}}(\alpha_k)$ in 
\eqref{eq_atofB}, the belief of $\mathbf H_k$ at factor node $f_k^B$ is 
Bernoulli Gaussian and expressed as 
\begin{align} 
\label{eq_mfkB}
\mathcal{M}_{f_k^B}(\mathbf H_k) & \propto \sum_{\alpha_k 
= 0}^1  p(\mathbf H_k|\alpha_k) \mathcal M_{\alpha_k \to 
f^B_{k}}(\alpha_k) \mathcal{M}_{\mathbf H_k \to f_k^B}(\mathbf H_k) 
\notag \\ 
& = (1-\lambda_{k}^{B,post})\delta(\mathbf{H}_k) + \lambda_{k}^{B,post} 
{\cal CN}(\mathbf{H}_k; \boldsymbol \mu_k^B ; \mathbf \Phi_k^B)
\end{align}
where
\begin{align}
	 & \lambda_{k}^{B,post}  = \left(1 + \frac{ (1-\lambda_{k}^{B,pri}) 
	 \cdot {\cal CN}(\mathbf 0 ; \mathbf H_k^{B,pri} ,\mathbf V^B )} 
	 {\lambda_{k}^{B,pri} \cdot {\cal CN}(\mathbf 0 ; \mathbf H_k^{B,pri}, 
	 \mathbf V^B + \vartheta_{\mathbf H} \mathbf I ) }\right)^{-1} \notag \\
	 &\mathbf \Phi_k^B = (\vartheta_{\mathbf H}^{-1} \mathbf I + 
	 (\mathbf  V^B)^{-1})^{-1} \notag \\
	 &\boldsymbol \mu_k^B  = {\vartheta_{\mathbf H}} 
	 {(\vartheta_{\mathbf H}{\bf{I}} + \mathbf V^B)^{-1}} \cdot {\rm 
	 vec}(\mathbf H_{k}^{B,pri}).
\end{align}
We note that the mean of ${\rm vec}(\mathbf H_k)$ with respect to 
$\mathcal{M}_{f_k^B}(\mathbf H_k)$ is
\begin{align}
	{\rm vec}(\mathbf H_k^{B,post}) = \lambda_{k}^{B,post} \boldsymbol \mu_k^B. 
\end{align}
Define $l = (m-1)Q + q$. The variance of the $(q,m)$-th element of 
$\mathbf H_k$  with respect to $\mathcal{M}_{f_k^B}(\mathbf H_k)$ is
\begin{equation}
\vartheta_{h_{k,q,m}}^{B,post} = \lambda_{k}^{B,post} \left( |\mu_{k,l}^B|^2 +  
\Phi_{k,l,l}^B  \right) - |h_{k,q,m}^{B,post}|^2 .  
\end{equation}

Recall that the relationship between $\mathbf H_k$ and $\mathbf h_m$ is 
described as $[\mathbf H_1^T,...,\mathbf H_K^T]^T = [\mathbf h_1, ..., 
\mathbf h_M]$. Through such relationship, we obtain $\mathbf 
h_m^{B,post}$. Following \cite{Ma_1,TARM}, the variance 
${v}_{\mathbf h_m}^{B,post}$ is approximated by
\begin{equation}
{v}_{\mathbf h_m}^{B,post} = \frac{1}{KQ} \sum_{k,q} 
\vartheta_{h_{k,q,m}}^{B,post}.
\end{equation}
Then the belief of $\mathbf h_m$ at Module B is defined as
\begin{equation} \label{Bm}
\mathcal{M}_B (\mathbf h_m) \sim \mathcal{CN}(\mathbf h_m ; \mathbf h_m^{B,post} , {v}_{\mathbf h_m}^{B,post} \mathbf I ).
\end{equation}
The above Gaussian belief assumption is widely used in message passing 
based iterative algorithms such as Turbo-CS \cite{Ma_1}, approximate 
message passing (AMP) \cite{AMP_1} and expectation propagation (EP) 
\cite{EP}. Such treatment may loses some information but facilities the 
message updates. (Strictly speaking, the belief of the variable corresponds to 
a factor node instead of a Module. However, the belief of $\mathbf h_m$ is 
associated with $K$ factor nodes $p(\mathbf H_k | \alpha_k), \forall k$, 
making the expression cumbersome. For convenience, we use the definition 
$\mathcal{M}_B (\mathbf h_m)$.)

From the sum-product rule, we have $ \mathcal{M}_{\mathbf h_m 
\to B}(\mathbf h_m) = {\mathcal{M}_{\mathbf y_m \to \mathbf 
h_m}(\mathbf h_m)}$. Then the extrinsic message is calculated as 
\begin{equation}
\mathcal{M}_{ B \to \mathbf h_m}(\mathbf h_m) 
\propto \frac{\mathcal{M}_{B}(\mathbf h_{m})}{\mathcal{M}_{\mathbf 
y_m \to \mathbf h_m}(\mathbf h_m)}.
\end{equation}
Given the Gaussian messages $\mathcal{M}_B(\mathbf h_{m})$ and 
$\mathcal{M}_{\mathbf h_m \to B(\mathbf h_{m})}$, the extrinsic 
message is also Gaussian with its variance and mean respectively given by
\begin{align}
& {v}_{\mathbf h_m}^{B,ext} =
\left(\frac{1}{{v}_{\mathbf h_m}^{B,post}}-\frac{1}{{v}_{\mathbf 
h_m}^{B,pri}}\right)^{-1} \notag \\
& \mathbf{h}_{m}^{B,ext} ={v}_{\mathbf h_m}^{B,ext} 
\left(\frac{\mathbf{h}_{m}^{B,post}} {{v}_{\mathbf h_m}^{B,post}} 
-\frac{\mathbf{h}_{m}^{B,pri}}{{v}_{\mathbf h_m}^{B,pri}}\right)
\end{align}
where $\mathbf{h}_{m}^{B,ext}$ and ${v}_{\mathbf h_m}^{B,ext}$ are 
respectively used as the input mean and variance of $\mathbf{h}_{m}$ for 
Module A, i.e., $\mathbf{h}_{m}^{A,pri} = \mathbf{h}_{m}^{B,ext}$ and 
${v}_{\mathbf h_m}^{A,pri} = {v}_{\mathbf h_m}^{B,ext}$. Then we 
have 
\begin{equation}
\mathcal{M}_{\mathbf h_m \to \mathbf y_m}(\mathbf h_m) = \mathcal{M}_{ B \to \mathbf h_m}(\mathbf h_m) \sim \mathcal{CN}(\mathbf h_m ; \mathbf h_m^{A,pri},v_{\mathbf h_m}^{A,pri} \mathbf I)
\end{equation}

\subsection{Module C: Denoiser of $\mathbf C_k$}
Similarly to the process in Module B, each $\mathbf C_k, \forall k$ in 
module C is estimated individually by exploiting the prior $ p(\mathbf 
C_k|\alpha_k)$ and the messages from Modules A and D. For description 
convenience, $f_k^C$ is used to replace factor node $p(\mathbf C_k | 
\alpha_k)$.  The message from variable node $\mathbf C_k$ to factor node 
$p(\mathbf C_k | \alpha_k)$ is expressed as
\begin{align}
\mathcal{M}_{\mathbf C_k \to f_k^C}(\mathbf C_k)  \sim {\cal{CN}} 
(\mathbf C_k ; \mathbf C_k^{C,pri} ,  \mathbf V^C )  
\end{align}
where $\mathbf C_k^{C,pri} \in \mathbb C^{Q\times M}$ with its 
$(q,m)$-th element $c_{k,q,m}^{C,pri} = c_{k,q,m}^{A,ext}$ and 
$\mathbf V^C = {\rm diag}([v_{\mathbf c_1}^{C,pri},...,v_{\mathbf 
c_M}^{C,pri}]^T \otimes \mathbf 1_Q^T )$. 

With the Bernoulli Gaussian prior $ p(\mathbf C_k|\alpha_k)$, 
Gaussian message $\mathcal{M}_{\mathbf C_k \to f_k^C}(\mathbf C_k)$ 
and message $\mathcal{M}_{\alpha_k \to f^C_{k}}(\alpha_k)$ in 
\eqref{eq_BtoC}, the belief of $\mathbf C_k$ at factor node $f_k^C$ is 
Bernoulli Gaussian and expressed as 
\begin{align} 
\mathcal{M}_{f_k^C}(\mathbf C_k) & \propto \sum_{\alpha_k 
=0}^1 p(\mathbf C_k|\alpha_k) \mathcal M_{\alpha_k \to 
f^C_{k}}(\alpha_k) \mathcal{M}_{\mathbf C_k \to f_k^C}(\mathbf C_k) 
\notag \\ 
& = (1-\lambda_{k}^{C,post})\delta(\mathbf{C}_k) + \lambda_{k}^{C,post} 
{\cal CN}(\mathbf{C}_k; \boldsymbol \mu_k^C ; \mathbf \Phi_k^C)
\end{align}
where
\begin{align}
& \lambda_{k}^{C,post}  =  \left(1 + \frac{ (1-\lambda_{k}^{C,pri}) \cdot 
{\cal CN}(\mathbf 0 ; \mathbf C_k^{C,pri} ,\mathbf V^C )} 
{\lambda_{k}^{C,pri} \cdot {\cal CN}(\mathbf 0 ; \mathbf C_k^{C,pri}, 
\mathbf V^C + \vartheta_{\mathbf C} \mathbf I ) }\right)^{-1} \notag \\
&\mathbf \Phi_k^C = (\vartheta_{\mathbf C}^{-1} \mathbf I + (\mathbf 
V^C)^{-1})^{-1} \notag \\
&\boldsymbol \mu_k^C  = {\vartheta_{\mathbf C}} {(\vartheta_{\mathbf 
C}{\bf{I}} + \mathbf V^C)^{-1}} \cdot {\rm vec}(\mathbf C_{k}^{C,pri}).
\end{align}
Then the mean of $\mathbf C_k$ with respect to 
$\mathcal{M}_{f_k^C}(\mathbf C_k)$ is
\begin{align}
{\rm vec}(\mathbf C_k^{post}) = \lambda_{k}^{C,post} \boldsymbol \mu_k^C. 
\end{align}
The variance of the $(q,m)$-th element of $\mathbf C_k$ is given by
\begin{equation}
\vartheta_{c_{k,q,m}}^{C,post} = \lambda_{k}^{C,post} \left( |\mu_{k,l}^C|^2 +  
\Phi_{k,l,l}^C  \right) - |c_{k,q,m}^{C,post}|^2.
\end{equation}
Similarly to \eqref{Bm}, we define the belief of $\mathbf c_m$ at Module C 
as
\begin{equation}
\mathcal{M}_{C}(\mathbf c_m) \sim \mathcal{CN}(\mathbf c_m ; \mathbf 
c_m^{C,post} , {v}_{\mathbf c_m}^{C,post} \mathbf I )
\end{equation}
with the variance ${v}_{\mathbf c_m}^{C,post}$ given by
\begin{equation}
{v}_{\mathbf c_m}^{C,post} = \frac{1}{KQ} \sum_{k,q} 
\vartheta_{c_{k,q,m}}^{C,post}.
\end{equation}

Then we calculate the Gaussian extrinsic message with its variance and mean 
as follows :
\begin{align}
& {v}_{\mathbf c_m}^{C,ext} =
\left(\frac{1}{{v}_{\mathbf c_m}^{C,post}}-\frac{1}{{v}_{\mathbf 
c_m}^{C,pri}}\right)^{-1} \notag  \\
& \mathbf{c}_{m}^{A,pri} = {v}_{\mathbf c_m}^{C,ext}
\left(\frac{\mathbf{c}_{m}^{C,post}} {{v}_{\mathbf c_m}^{C,post}} 
-\frac{\mathbf{c}_{m}^{C,pri}}{{v}_{\mathbf c_m}^{C,pri}}\right).
\end{align}
The input mean and variance of $\mathbf c_m$ for module A are 
respectively set as $\mathbf{c}_{m}^{A,pri} = \mathbf{c}_{m}^{C,ext}$ 
and ${v}_{\mathbf c_m}^{A,pri} = {v}_{\mathbf c_m}^{C,ext}$.

\subsection{Module D: Estimation of $\alpha_k$}
In Module D, we calculate the messages $\mathcal M_{\alpha_k \to 
f^B_{k}}(\alpha_k)$ and $\mathcal M_{\alpha_k \to f^C_{k}}(\alpha_k)$ 
as the inputs of Modules B and C, respectively. Furthermore, the device 
activity is detected by combing the messages from Modules B and C. 
According to the sum-product rule, the message from factor node 
$f_{k}^B$ to variable node $\alpha_k$ is
\begin{align} \label{ftoa}
\mathcal M_{ f_{k}^B \to \alpha_k }(\alpha_k) & \propto \int_{\mathbf H_k} 
f_{k}^B (\mathbf H_k,\alpha_k) \cdot \mathcal M_{\mathbf H_k \to f_{k}^B} 
(\mathbf H_k) \notag \\ 
	& = (1- \pi_k^B )\delta[\alpha_k]+\pi_k^B \delta[1-\alpha_k]
\end{align}
where 
\begin{equation}
\pi_k^B = \left(1 + \frac{{\cal CN}(\mathbf 0;\mathbf H_k^{B,pri},\mathbf V^B) 
} {{\cal CN}(\mathbf 0;\mathbf H_k^{B,pri}, \mathbf V^B + \vartheta_{\mathbf 
H} \mathbf I)} \right)^{-1}.
\end{equation}
Then the message from variable node $\alpha_k$ to factor node $f_k^C$ is 
\begin{align} \label{eq_BtoC}
	\mathcal M_{\alpha_k \to f_k^C} (\alpha_k) & \propto \mathcal M_{ f^B_{k} 
	\to \alpha_k } \cdot p(\alpha_k) \notag \\ & =  
	(1-\lambda_{k}^{C,pri})\delta[\alpha_k]+\lambda_{k}^{C,pri} 
	\delta[1-\alpha_k]
\end{align}
with
\begin{equation} \label{eq_l_Bext}
\lambda_{k}^{C,pri} = \frac{\lambda \cdot \pi_k^B 
}{\lambda \cdot \pi_k^B  + (1-\lambda) \cdot (1-\pi_k^B)}.
\end{equation}

Similarly to \eqref{ftoa}, the message from factor node $f_{k}^C$ to 
variable node $\alpha_k$ is 
\begin{align} \label{eq_lC}
\mathcal M_{ f_{k}^C \to \alpha_k } = (1- \pi_k^C 
)\delta[\alpha_k]+\pi_k^C 
\delta[1-\alpha_k]
\end{align}
where 
\begin{equation} \label{eq_piC}
\pi_k^C = \left(1 + \frac{{\cal CN}(\mathbf 0;\mathbf C_k^{C,pri},\mathbf V^C) 
} {{\cal CN}(\mathbf 0;\mathbf C_k^{C,pri}, \mathbf V^C + \vartheta_{\mathbf 
C} \mathbf I)} \right)^{-1}.
\end{equation}
Then the message from variable node $\alpha_k$ to factor node $f^B_{k}$ 
is 
\begin{align} \label{eq_atofB}
\mathcal M_{\alpha_k \to f^B_{k}}(\alpha_k) & \propto \mathcal M_{ f^C_{k} \to 
\alpha_k } \cdot p(\alpha_k) \notag \\
& = (1-\lambda_{k}^{B,pri})\delta[\alpha_k]+\lambda_{k}^{B,pri} 
\delta[1-\alpha_k]
\end{align}
with 
\begin{equation}
	\lambda_{k}^{B,pri} =  \frac{\lambda \cdot \pi_{k}^{C} } {\lambda \cdot 
	\pi_{k}^{C}  + (1-\lambda) \cdot (1 - \pi_{k}^{C}) }.
\end{equation}

Define the belief of $\alpha_k$ at factor node $p( \alpha_k )$ as $\mathcal 
M_k(\alpha_k)$. From the sum-product rule, $\mathcal M_k(\alpha_k)$ is 
given by 
\begin{align}
\mathcal M_k(\alpha_k) & \propto \mathcal M_{ f_{k}^B \to \alpha_k }(\alpha_k) \mathcal M_{ f_{k}^C \to \alpha_k }(\alpha_k) p(\alpha_k) \notag \\
& = (1-\lambda_{k}^{D,post})\delta[\alpha_k]+\lambda_{k}^{D,post} 
\delta[1-\alpha_k]
\end{align}
where
\begin{equation}
	\lambda_{k}^{D,post} =  \frac{\lambda \cdot \pi_{k}^{B} \cdot \pi_{k}^{C} } 
	{\lambda \cdot \pi_{k}^{B} \cdot \pi_{k}^{C}  + (1-\lambda)  \cdot (1 - 
	\pi_{k}^{B}) \cdot (1 - \pi_{k}^{C}) }.
\end{equation}
Clearly, $\lambda_k^{D,post}$ ($0 \leq \lambda_k^{D,post} \leq 1$) 
indicates the probability that the $k$-th device is active. Therefore, we adopt 
a threshold-based strategy for device activity detection as
\begin{equation}
\hat \alpha_{k}=\left\{\begin{array}{l}
1, \quad \lambda_k^{D,post} \geq \lambda^{thr} \vspace{3pt} \\
0, \quad \lambda_k^{D,post} < \lambda^{thr} 
\end{array} \quad k=1,...,K \right.
\end{equation}
where $\lambda^{thr}$ is a predetermined threshold. 

% Note that the power of channel compensation matrix $\mathbf C_k$ is the minority of the power of $\mathbf G_k$. In other word, the power of $\mathbf G_k$ is dominated by channel mean matrix $\mathbf H_k$. It means that the estimate of the message $\mathcal M_{ f_{k}^C \to \alpha_k }(\alpha_k)$ is less accurate than that of $\mathcal M_{ f_{k}^B \to \alpha_k }(\alpha_k)$. In practice, when the system work in a relative low signal-to-noise ratio (SNR), the calculation of $\mathcal M_k(\alpha_k)$ can ignore the message $\mathcal M_{ f_{k}^C \to \alpha_k }(\alpha_k)$.

\subsection{Pilot Design and Complexity Analysis} \label{pilot}
The overall algorithm is summarized in Algorithm \ref{alg1}. In each 
iteration, the channel mean matrix $\mathbf H$ is first updated (step 2-8) and then the channel compensation matrix $\mathbf C$ is updated (step 9-15). This is because the power of the channel mean matrix $\mathbf H$ is dominant and the iteration process effectively suppresses error propagation. Note that the prior parameters learning is shown in the next section.

As mentioned in Section IV-B, the pilot matrix $\mathbf A$ is required to be partial orthogonal. To fulfill this requirement, the pilot symbols $\{a_{k,n}^{(t)}\}_{k=1}^K$ transmitted on the $n$-th subcarrier at the $t$-th OFDM symbol has the following property:
\begin{equation}
[a_{1,n}^{(t)},...,a_{k,n}^{(t)},...,a_{K,n}^{(t)}] = \mathbf u_i
\end{equation}
where $\mathbf u_i$ is a row vector randomly selected from an orthogonal matrix $\mathbf U \in \mathbb C^{K \times K}$ and the selected row is different for different $n,t$. Combing $\mathbf A = [\mathbf \Lambda_1 \mathbf E_1,...,\mathbf \Lambda_K \mathbf E_1]$ and the definition of $\mathbf \Lambda_k$ in \eqref{Lambda}, it is easy to verify the partial orthogonality of $\mathbf A$.

We next show that when $\mathbf U$ is the discrete Fourier transform (DFT) matrix, the algorithm complexity can be further reduced. To enable the fast Fourier transform (FFT), the pilot matrix $\mathbf A$ is expressed as
\begin{equation} \label{fast}
	\mathbf A = {\rm diag (\mathbf S_1 \mathbf U,...,\mathbf S_Q \mathbf U)} \mathbf P 
\end{equation} 
where $\mathbf S_q \in \mathbb R^{TN/Q \times K}$ is a row selection 
matrix consisting of $TN/Q$ randomly selected rows from the $K \times K$ 
identity matrix. (The selected rows are different for different $\mathbf S_q$.) 
$\mathbf P = [\mathbf P_1,...,\mathbf P_K]$ is a column exchange matrix. 
In specific, in the $q$-th column of $\mathbf P_k \in \mathbb R^{KQ \times 
Q},\forall k$, only the $k+K(q-1)$-th row is one while the others are zeros. 
Note that $|a_{k,n}^{(t)}|^2=P$ and matrix $\mathbf B$ has the same fast transform as $\mathbf A$ due to the relationship $\mathbf B = \mathbf D \mathbf A$.

By using the FFT, the estimates of $ \mathbf h_m \in \mathbb C^{QK} $ and $ \mathbf c_m \in \mathbb C^{QK}, \forall m$ in Module A has the complexity ${\cal O}(QMK {\rm log_2} K)$. In modules B, C and D, the estimates of $  \mathbf H_k \in \mathbb C^{Q \times M}$, $ \mathbf C_k \in \mathbb C^{Q \times M}$, and $\alpha_k, \forall k,$ involve vector multiplications with the complexity ${\cal O}(QMK)$. As a result, the overall complexity of Turbo-MP is ${\cal O}(QMK {\rm log_2} K) + {\cal O}(QMK)$ per iteration. It is worth noting that the algorithm complexity is linear to the antenna number $M$ and is approximately linear to the device number $K$.

\begin{algorithm}
 \caption{\label{alg1}Turbo Message Passing (Turbo-MP)}
 \begin{algorithmic}
  
  \REQUIRE $\mathbf{Y}$, $\mathbf A$, $\mathbf B$.
%  received signal $\mathbf{Y}_{f} = [\mathbf{y}^{(1)}_{f}, 
%  \cdots, 
%  \mathbf{y}^{(P)}_{f}]$, pilot matrix $\mathbf{X}^{(p)}$ $\forall p$, 
%and 
%  additive noise variance $\sigma^{2}$.
  
 \quad \ 1: Initialize $\boldsymbol{\theta}$ by the EM or NN approach.
 
 \textbf{while} the stopping criterion is not met  \textbf{do}
 
\textbf{\% Module A: Linear estimation of  $\mathbf h_m$}
  
 \quad \ 2: Update $\mathbf h_m^{A,post}$, $v_{\mathbf h_m}^{A,post}$ 
 by (19) and (21)-(22), $\forall m$.
  
 \quad \  3: Update  $\mathbf h_m^{A,ext}$, $v_{\mathbf h_m}^{A,ext}$ 
 by (25), $\forall m$.
  
\textbf{\% Module B: Denoiser of  $\mathbf H_k$}
  
 \quad \  4: Update $\mathbf H_k^{B,pri}$, $\mathbf V^B$ by (29), $\forall 
 k$.
 
 \quad \  5: Update $\mathcal{M}_{f_k^C \to \alpha_k}$, $\mathcal{M}_{ 
 \alpha_k \to f_k^B}$ by (51)-(54), $\forall k$.
 
 \quad \  6: Update $\mathbf H_k^{B,post}$, $v_{h_{k,q,m}}^{B,post}$ by 
 (30)-(33), $\forall k$.
 
  \quad \  7: Update $\mathbf h_m^{B,post}$, $v_{\mathbf h_m}^{B,post}$ 
  by (34), $\forall m$.
  
 \quad \  8: Update $\mathbf h_m^{B,ext}$, $v_{\mathbf h_m}^{B,ext}$ 
   by (37), $\forall m$.
  
\textbf{\% Module A: Linear estimation of  $\mathbf c_m$}

 \quad \ 9: Update $\mathbf c_m^{A,post}$, $v_{\mathbf c_m}^{A,post}$ 
 by (21) and (26), $\forall m$.
  
 \quad \  10: Update  $\mathbf c_m^{A,ext}$, $v_{\mathbf c_m}^{A,ext}$ 
 by (27), $\forall m$.
 
\textbf{\% Module C: Denoiser of  $\mathbf C_k$}
  
 \quad \  11: Update $\mathbf C_k^{C,pri}$, $\mathbf V^C$ by (39), 
 $\forall k$.
 
 \quad \  12: Update $\mathcal{M}_{f_k^B \to \alpha_k}$, $\mathcal{M}_{ 
 \alpha_k \to f_k^C}$ by (47)-(50), $\forall k$.
 
 \quad \  13: Update $\mathbf C_k^{C,post}$, $v_{c_{k,q,m}}^{C,post}$ 
 by (40)-(43), $\forall k$.
 
  \quad \  14: Update $\mathbf c_m^{C,post}$, $v_{\mathbf 
  c_m}^{C,post}$ by (45), $\forall m$.
  
 \quad \  15: Update $\mathbf c_m^{C,ext}$, $v_{\mathbf c_m}^{C,ext}$ 
   by (46), $\forall m$.
  
\textbf{\% Parameters learning} 

 \quad \  16: Update $\boldsymbol{\theta}$ by EM or NN approach. 
 
 \textbf{end while}
 
 \quad \  17: Update $\lambda_{k}^{D,post}$ by (55) and (56), $\forall 
 k$.
 
 \ENSURE $\mathbf H_k^{B,post}$, $\mathbf C_k^{C,post}$, and 
 $\lambda_{k}^{D,post}$, $\forall k$. 
\end{algorithmic}
\end{algorithm}

\section{Parameters Learning}
The prior parameters $ \boldsymbol{\theta} = \{ \vartheta_{\mathbf H}, 
\vartheta_{\mathbf C}, \sigma_w^2 ,\lambda \} $ used in Turbo-MP are 
unknown and required to be estimated in practice. In the following, we utilize 
two machine learning methods, i.e., the EM and the NN approach, to 
learn these prior parameters.

\subsection{EM Approach}
We first use the EM algorithm \cite{EM} to learn $\boldsymbol 
\theta$. The EM process is described as
\begin{equation} \label{eq_EM}
\boldsymbol{\theta}^{(i+1)} =\arg \max_{\boldsymbol{\theta}} 
\mathbb{E}\left[\ln 
p(\mathbf{Y}, \mathbf{H},\mathbf{C},\boldsymbol{\alpha}  ; \boldsymbol 
\theta) \mid \mathbf{Y} ; 
\boldsymbol{\theta}^{(i)}\right]
\end{equation}
where $\boldsymbol \theta^{(i)}$ is the estimate of  $\boldsymbol \theta$ at  
the $i$-th EM iteration. $\mathbb{E}[\cdot | \mathbf{Y} ; 
\boldsymbol{\theta}^{(i)}]$ represents the expectation over the posterior 
distribution $p(\mathbf{H},\mathbf{C},\boldsymbol{\alpha} | \mathbf{Y} ; 
\boldsymbol{\theta}^{(i)})$. Note that it is difficult to obtain the true 
posterior distribution. Instead, we utilize the message products $\prod_k 
\mathcal{M}_{f_k^B}(\mathbf H_k) \mathcal{M}_{f_k^C}( 
\mathbf C_k ) \mathcal{M}_k (\alpha_k) $ as an approximation. Then we set 
the derivatives of $\mathbb{E}[\ln p(\mathbf{H},\mathbf{C}, 
\mathbf{Y} ; \boldsymbol \theta) \mid \mathbf{Y} ; 
\boldsymbol{\theta}^{(i)}]$ (with respect to  $\vartheta_{\mathbf H}$) 
to zero and obtain
\begin{align}
\vartheta_{\mathbf H}^{(i+1)} = \frac{\sum_k 
\lambda_k^{D,post} \left(|| \mathbf H_k^{B,post}||_F^2 + \sum_{q,m} 
\vartheta_{h_{k,q,m}}^{B,post} \right) }{QM \sum_k 
\lambda_k^{D,post}}.
\end{align}
Similarly, the EM estimate of $\vartheta_{\mathbf C}$ is given by 
\begin{align}
\vartheta_{\mathbf C}^{(i+1)} = \frac{\sum_k 
\lambda_k^{D,post} \left(|| \mathbf C_k^{C,post}||_F^2 + \sum_{q,m} 
\vartheta_{c_{k,q,m}}^{C,post} \right) }{QM \sum_k 
\lambda_k^{D,post}}.
\end{align}
The EM estimate of  $\sigma_w^2$ is given by 
\begin{align}
\label{EM_w}
 (\sigma_w^2)^{(i+1)} = & \frac{1}{MTN} \Big |\Big|\mathbf Y - \mathbf 
 A \mathbf H^{B,post} -   \mathbf B \mathbf C^{C,post} \Big 
 |\Big|_F^2	\notag \\ 
& + \frac{K}{M} \sum_{m}\big ( v_{\mathbf h_m}^{B,post} 
    + \frac{1}{TN} \sum_{i=1}^{TN} D_{i,i}^2 v_{\mathbf 
    c_m}^{C,post} \big ).
\end{align}
The EM estimate of $\lambda$ is given by 
\begin{equation}
	\lambda^{(i+1)} = \frac{1}{K}\sum_k \lambda_k^{D,post}.
\end{equation}

%As an iterative algorithm of parameter learning, EM algorithm may 
%converge 
%to a local maximum of the likelihood function. Therefore, a proper 
%initialization of the prior parameters is crucial. Assume that the path loss 
%$\rho_k$ and signal power $P_k$ are acquired at BS, the initialization of 
%the 
%prior parameters is given by 
%\begin{align} \label{EM_w}
%& \vartheta_{\mathbf H}^{(0)}  = \frac{1}{K} \sum_k P_k \rho_k \notag 
%\\
%& \vartheta_{\mathbf C}^{(0)}  = \frac{PT} {Q ||\mathbf d||_2^2 } 
%\cdot \gamma  \vartheta_{\mathbf H}^{(0)} \notag \\
%& (\sigma_w^2)^{(0)} = \frac{1}{MTN} || \mathbf Y ||_F^2
%\end{align}
%where the use of  $ \gamma $  is because $\vartheta_{\mathbf C}$ is 
%partly influenced by the severity of the frequency-selectivity. From the 
%simulation in various multi-path channels, $\gamma \in [0.01,0.05]$ is a 
%suitable scope.
Turbo-MP algorithm with EM approach to learn $\boldsymbol{\theta}$ is shown in Algorithm 1, which we refer to as Turbo-MP-EM. In practice, we can update $\mathbf H$ several times and then update $\mathbf C$ once to improve the algorithm stability. Besides, it is known the estimation accuracy of $\vartheta_{\mathbf H}$, $\vartheta_{\mathbf C}$, and $\lambda$ by EM relies on the accuracy of the posterior approximation in Turbo-MP, and inaccurate estimation may affect the algorithm convergence. Therefore, we recommend to update $\vartheta_{\mathbf H}$, $\vartheta_{\mathbf C}$, and $\lambda$ once after $\mathbf H$ and $\mathbf C$ is updated several times. In the first algorithm iteration, we initialize $\vartheta_{\mathbf H}^{(0)}=1$, $\vartheta_{\mathbf C}^{(0)}=10^{-3}$, and $\lambda^{(0)}=0.1$. As for the update of $\sigma_w^2$, it is updated in each iteration of Turbo-MP with the initialization $(\sigma_w^2)^{(0)}\!\! = \!\! ||\mathbf Y||_F^2/MTN$. However, we find the second part of \eqref{EM_w}, i.e., $\frac{K}{M} \sum_{m}\big ( v_{\mathbf h_m}^{B,post} + \frac{1}{TN} \sum_{i=1}^{TN} D_{i,i}^2 v_{\mathbf c_m}^{C,post} \big )$ is negligible in most cases while in worst case, it continues to rise with the increase of Turbo-MP iterations. Therefore, we delete the second part of \eqref{EM_w} in simulation.

\subsection{NN Approach}
% An alternative strategy is to learn the prior parameters from data, which 
% improves the robustness of the algorithm in real implementation. 

In deep learning [3], training data can be used to train the parameters of a 
deep neural network. Inspired by the idea in \cite{LAMP} and \cite{LTMP}, 
we unfold the iterations of Turbo-MP algorithm and regard it as a 
feed-forward NN termed Turbo-MP-NN. In specific, each iteration represents one 
layer of the feed-forward NN where 
$\boldsymbol{\theta}$ is seen as the network parameter. We hope that with an appropriately defined loss function, Turbo-MP-NN can adaptively learn $\boldsymbol \theta$ from the training data. 

\begin{figure}[t]
	\centering
	\includegraphics[width=3.4in]{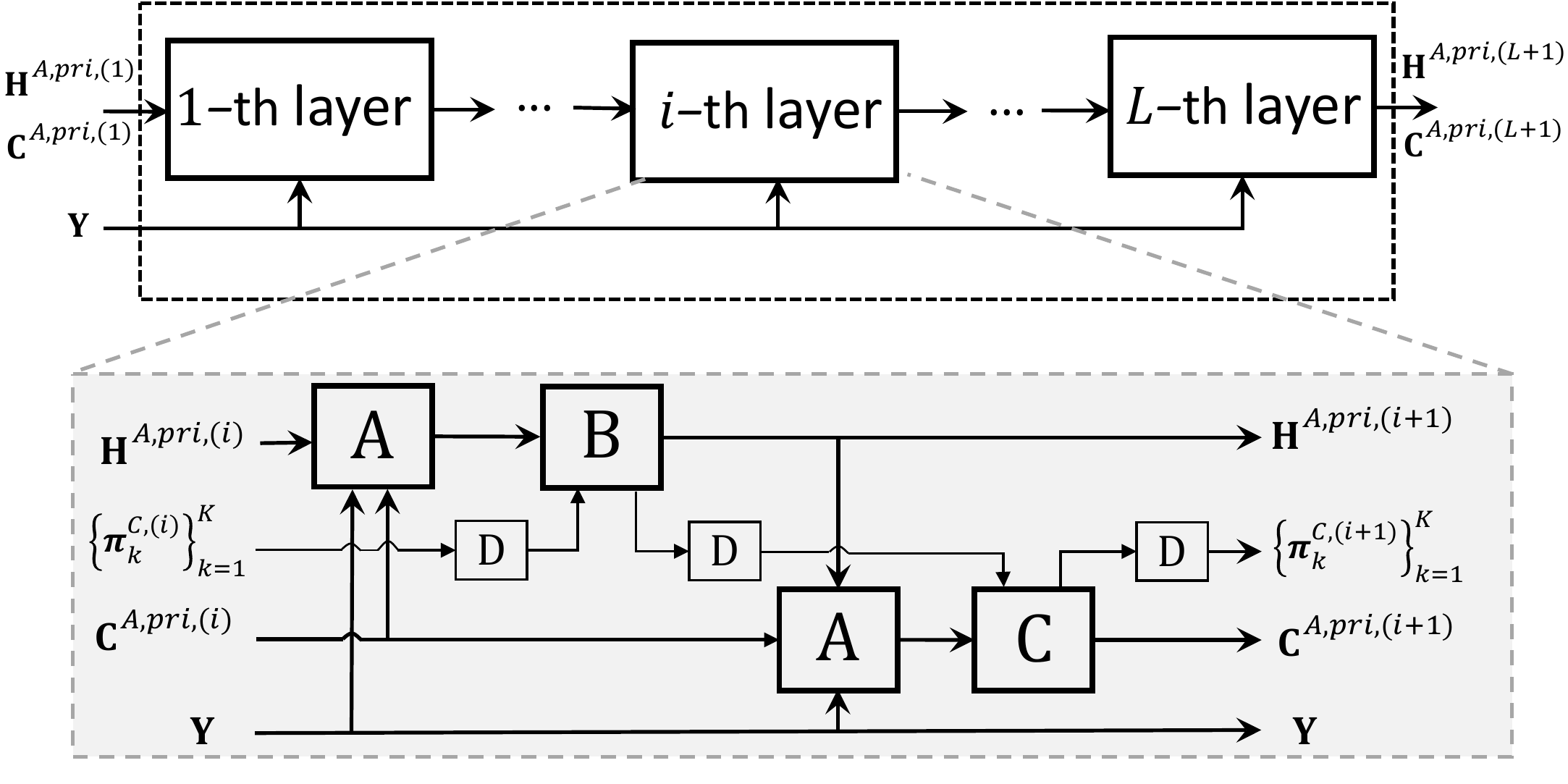}
	\caption{The block diagram of Turbo-MP-NN. The iterations of Turbo-MP are 
	unfolded into a neural network.	Capital letters A, B, C, D denote modules A, B, 
	C, and D, respectively.
%	The network consists of $L$ 
%	layers, and each layer has the same structure that contains two Module A, 
%	one Module B, one Module C and three module D. $\mathbf{{H}}^{A, pri, 
%(i)}$ and $\mathbf{{C}}^{A, pri, (i)}$ are the extrinsic messages passing to 
%the 
%$i$-th layer, while the estimated $\mathbf{\hat{H}}^{(i)}$ and 
%$\mathbf{\hat{C}}^{ (i)}$generated by module B and module C are not shown 
%in this figure.
	}
	\label{LMP}
\end{figure}

The structure of Turbo-MP-NN is shown in Fig. \ref{LMP} which consists of $L$ layers. Each layer has the same structure, i.e., 
the same linear and non-linear operations following step 2-15 in algorithm 1. 
To distinguish the estimates at different layer, denote $\mathbf H^{A,pri}$ 
,$\mathbf C^{A,pri}$ and $\pi_k^{C}$ obtained at the $i-1$-th layer 
(iteration) by  $\mathbf H^{A,pri,(i)}$, $\mathbf C^{A,pri,(i)}$, and 
$\pi_k^{C,(i)}$, respectively. In the $i$-th layer, the inputs consist of 
training data $\mathbf Y$ and the outputs of the $i-1$-th layer including 
$\mathbf H^{A,pri,(i)}$, $\mathbf C^{A,pri,(i)}$, and 
$\{\pi_k^{C,(i)}\}_{k=1}^K$. The loss function is defined as the normalized 
mean square error (NMSE) of channel estimation given by

\begin{align}
f(\boldsymbol{\theta})=\sum_k   \frac{  || \mathbf{G}_k - 
\mathbf{E}_1 \mathbf{\hat H}_k^{(L)} - 	
\mathbf{E}_2 \mathbf{\hat C}_k^{(L)} ||_F^2  }   { || 
\mathbf{G}_k ||_F^2  },
\end{align}
where the estimates $\mathbf{\hat H}_k^{(L)}$ and $ \mathbf{\hat C}_k^{(L)}$ can 
be $\mathbf H_k^{B,post}$ and $\mathbf C_k^{C,post}$ or $\mathbf H_k^{A,pri}$ 
and $\mathbf C_k^{C,pri}$ obtained in the $L$-th layer. Note that different from the 
EM approach,  all layers of the NN have the same $\boldsymbol{\theta}$. To avoid 
over-fitting, we train the neural network through the layer-by-layer method 
\cite{LTMP}. Specifically, we begins from the training of the first layer, then first two 
layers, and finally 
$L$ layers.  $\boldsymbol{\theta}$ is initialized following (63). For the first 
$i$ layers $i=1,2,\ldots,L$, we optimize $\boldsymbol{\theta}$ by using the 
back-propagation to minimize the loss function
\begin{align}
	f^{(i)}(\boldsymbol{\theta})=\sum_k   \frac{  || \mathbf{G}_k - \mathbf{E}_1 \mathbf{\hat H}_k^{(i)} - \mathbf{E}_2 \mathbf{\hat C}_k^{(i)} ||_F^2  }   { || \mathbf{G}_k ||_F^2  }.
	\end{align}
The detailed training process is shown in Algorithm \ref{layer-by-layer}. 
Once trained, the iteration process of Turbo-MP-NN is illustrated in 
Algorithm 1. From the simulation, we find that for a fixed sub-block number 
$Q$, the change of the learned $\vartheta_{\mathbf H}$ and 
$\vartheta_{\mathbf C}$ is linear with respect to the signal-to-ratio 
$ {\rm SNR} =  P / \sigma_N^2 $, and the change of the 
$\sigma_w^2$ minus $\sigma_N^2$ is also linear with respect to the SNR. 
As such, there is no need to train the prior parameters offline at 
different SNR.
\begin{algorithm}[htb] 
		\caption{Parameter Training of Turbo-MP-NN via Layer-by-Layer 
		Method} 
		\label{layer-by-layer} 
		\begin{algorithmic}[1] 
		\REQUIRE ~~\\ 
		$\mathbf{Y}, \mathbf{A}$, $\mathbf B$.
		\ENSURE ~~\\ 
		\STATE Initialize $\boldsymbol{\theta}$.
		\label{ code:fram:extract }
		\FOR{$i \in [1,L]$}
		\STATE Run  $i$ iterations of Turbo-MP algorithm following  step 
		2-15 in Algorithm 1  to obtain $\widehat{\mathbf{H}}^{(i)}_k$ 
		and $\widehat{\mathbf{C}}^{(i)}_k$.
		\STATE With the loss function $f^{(i)}(\boldsymbol{\theta})$, use 
		back-propagation to update $\boldsymbol{\theta}$.
		\ENDFOR

		\label{code:fram:select}
		\RETURN $\boldsymbol{\theta}$. 
		\end{algorithmic}
\end{algorithm}

% With an appropriately defined loss function, the neural network can 
%adaptively learn the model parameters from the training data using 
%back-propagation.

\section{Numerical Results}
In this section, we evaluate the performance of the proposed Turbo-MP 
algorithm in channel estimation and device activity detection. The simulation 
setup is as follows. The BS is equipped with $M=4$ or $8$  antennas. 
$P=72$ OFDM subcarriers are allocated for the random access with 
subcarrier spacing $\Delta f = 15$ kHz. $K=1000$ devices access the BS 
and transmit their signals with probability $\lambda=0.05$ in each timeslot. 
We adopt the TDL-C channel model with 300 ns r.m.s delay spread and its 
detailed PDP can be found in TR 38.901 R14 \cite{TR38901}. For 
Turbo-MP-NN, we randomly generate $10^4$ channel realizations for 
training and the training data is divided into minibatches of size 4. To 
evaluate the performance of the proposed algorithm, we use the NMSE and 
the detection error probability ${\rm Pe}  = {\rm P_{miss}} +  {\rm 
P_{false}}$ as the performance metrics, where ${\rm P_{miss}} = 1/K 
\sum_k p(\hat \alpha_k=0|\alpha_k=1)$ is the probability of miss detection and ${\rm P_{false}} = 1/K \sum_k p(\hat \alpha_k=1|\alpha_k=0)$ is the probability of false alarm. Note that for the figures showing CE performance, each data point is averaged over $5000$ realizations. For the figures showing ADD performance, at each data point, the accumulative number of the detection errors is larger than $10^3$.

The baseline algorithms are as follows:
\begin{itemize}
\item \textbf{FD-GMMV-AMP}: The GMMV-AMP algorithm was proposed in 
\cite{Malong} to joint estimate the channel response on every subcarrier and detect 
the active devices based on the frequency-domain system model \eqref{eq_mt}. We 
assume that the prior parameters including the channel variance on every subcarrier, 
noise variance $\sigma_N^2$ and access probability $\lambda$ are known for 
FD-GMMV-AMP. (The following algorithms also use $\sigma_N^2$ and $\lambda$ 
as known parameters.) 

\item \textbf{TD-Gturbo-MMV}: We adopt the Gturbo-MMV algorithm \cite{Turbo_Liu} to achieve the ADD and CE based on the time-domain system model \eqref{eq_mt2}, which we refer to as TD-Gturbo-MMV. Note that the denoiser in \cite{Turbo_Liu} is extended and applied to $\tilde{ \mathbf  H}_k$. We further assume that the BS knows the delay spread of the time-domain channel, and thus the time-domain delay taps can be truncated to reduce the number of channel coefficients to be estimated. Specifically, there are $8$ delay taps left with $4$ taps at the head and $4$ taps at the tail which contains $99 \%$ energy of the time-domain channel. As a Bayesian algorithm, TD-Gturbo-MMV requires the PDP of the time-domain channel as prior information. One option is to assume that the exact PDP is known. However, in practice, the exact PDP of each device is difficult to acquire. Therefore, we also consider using a flat PDP with $8$ equal-power delay taps as the prior information.

\item \textbf{TD-Vector AMP}: The Vector AMP \cite{AMP_Liu} algorithm is adopted as another baseline in the time-domain system model \eqref{eq_mt2}. Similarly, we extend and apply the denoiser in \cite{AMP_Liu} to $\tilde{ \mathbf  H}_k$. TD-Vector AMP with the exact PDP and the flat PDP are both considered. 
\end{itemize}

\begin{figure}[h] 
\centering
\includegraphics[width=3.5in]{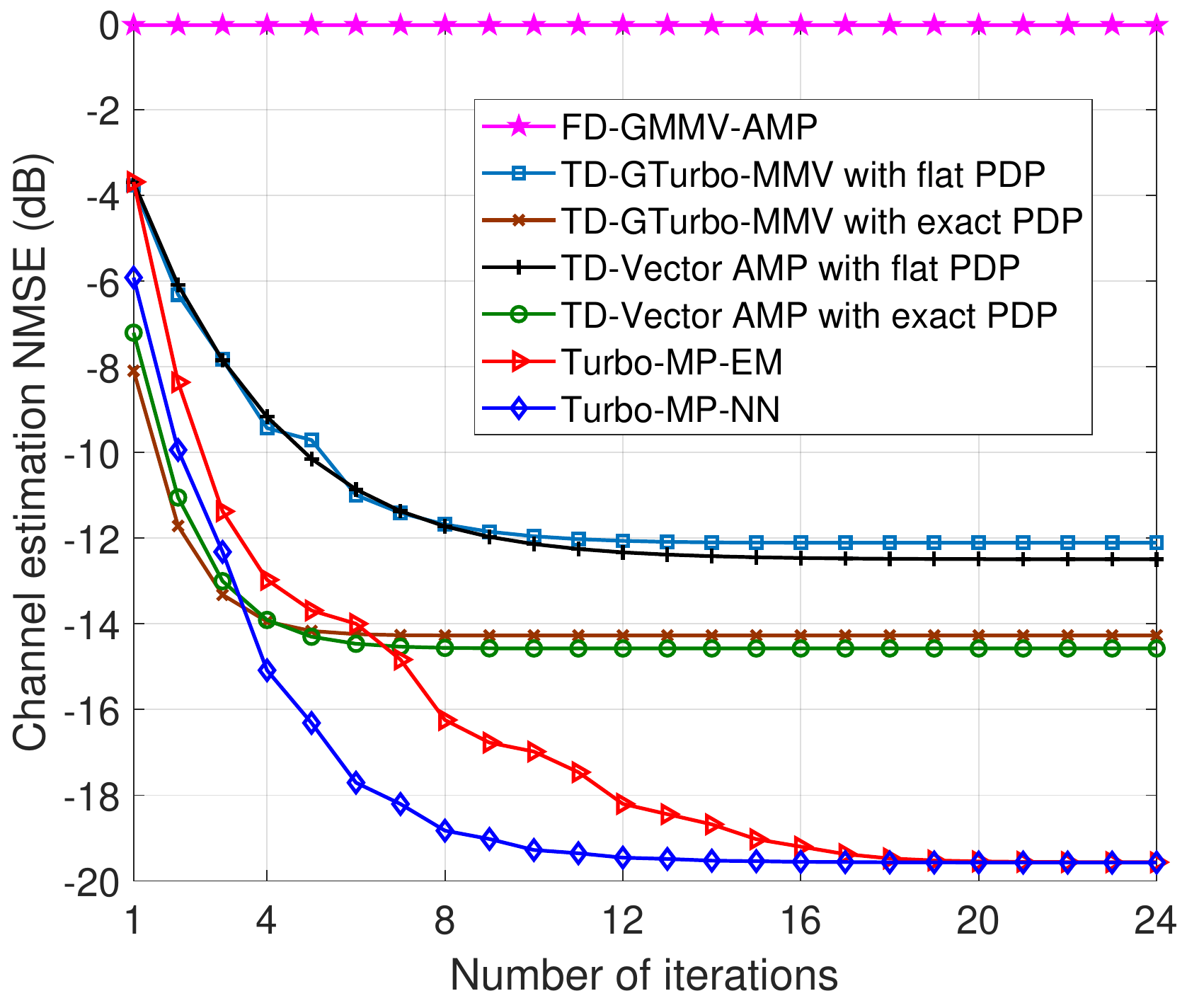}
\caption{Channel estimation NMSE versus Iterations number. The number 
of OFDM symbols $T=8$ and the sub-block number $Q=4$. ${\rm SNR}=10$ dB and the number of BS antennas $M=8$.}
\label{CE_1}
\end{figure}

Fig. \ref{CE_1} shows the channel estimation NMSE versus the iteration number. Both Turbo-MP-EM and Turbo-MP-NN converge and significantly outperform the other algorithms by more than $6$ dB in CE NMSE. Turbo-MP-NN converges faster than Turbo-MP-EM, which implying that the neural network approach can obtain more accurate prior parameters. In addition, it is seen that FD-GMMV-AMP has a quite poor performance since the average number of the unknown variables on every subcarrier is much larger than the number of the measurements, i.e., $\lambda K \gg T$. Concerning TD-GTurbo-MMV and TD-Vector AMP, there is a performance loss when the PDP is not exactly known.

\begin{figure}[h] 
	\centering
	\includegraphics[width=3.5in]{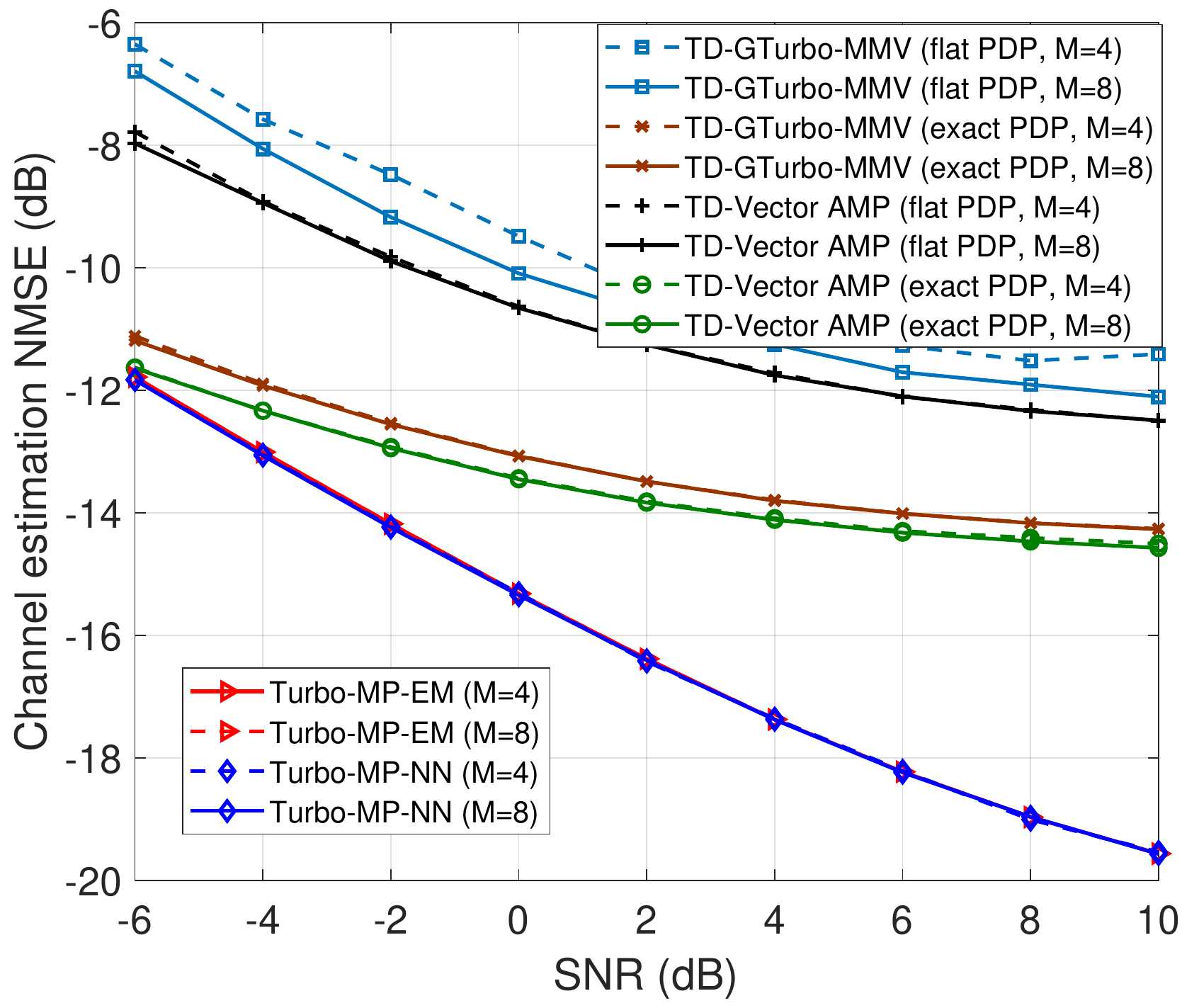}
	\caption{Channel estimation NMSE versus SNR. The number of OFDM 
	symbols $T=8$ and the sub-block number $Q=4$. }
\label{CE_2}
\end{figure}

\begin{figure}[h] 
	\centering
	\includegraphics[width=3.5in]{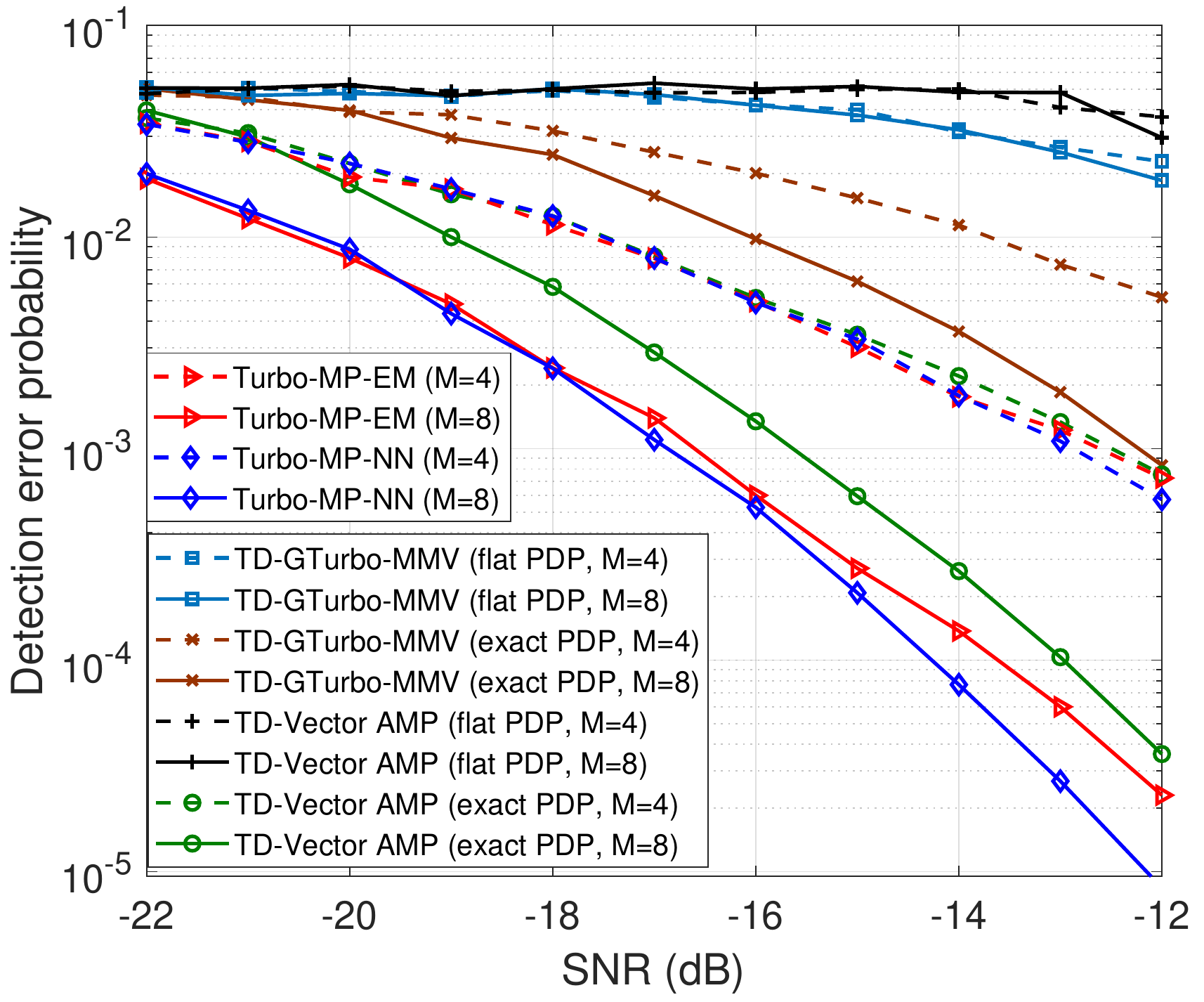}
	\caption{Detection error probability versus SNR. The number of OFDM symbols $T=8$ and the sub-block number $Q=4$. }
\label{CE_4}
\end{figure}

To further demonstrate the performance superiority of Turbo-MP, Fig. \ref{CE_2} shows the channel estimation NMSE against the SNR. With the increase of the SNR, the performance gap between Turbo-MP and the baselines becomes larger, which suggests that if the BS adopts the Turbo-MP algorithm to reach a high CE performance, the devices will consume much lower power. It is also found that the CE performance of each algorithm at different BS antenna number is similar. Fig. \ref{CE_4} shows the detection error probability versus the SNR. It is seen that as the antenna number $M$ increases, the detection performances of the tested algorithms improve more than one order of magnitude. This is because the increase of BS antennas leads to a larger dimension of the block-sparsity vector. Among the tested algorithms, Turbo-MP-NN has superiority over the other algorithms especially when antenna number $M=8$.

%the detection performance of Vector AMP with SCSI, Turbo-MP-EM and 
% is close. 

%Clearly, Vector AMP cannot work if the PDP is 
%not exact, which implies that the Bayesian detection algorithm of Vector 
%AMP highly relies on the exact prior. With exact PDP, the detection 
%performance of VAMP significantly improves. 

\begin{figure}[h] 
	\centering
	\includegraphics[width=3.5in]{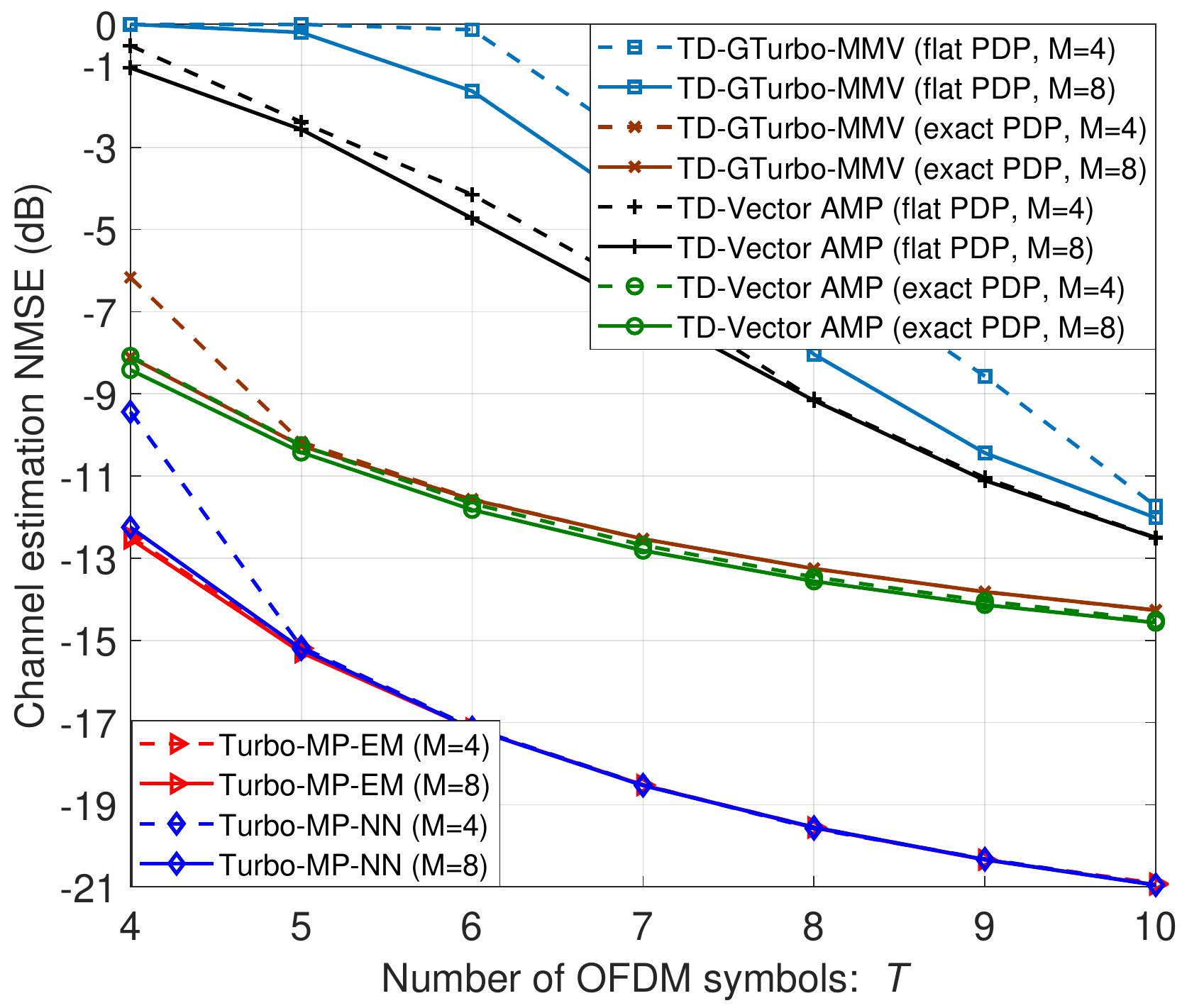}
	\caption{Channel estimation NMSE versus the number of OFDM symbols $T$. ${\rm SNR} = 10 $ dB and the sub-block number $Q=4$. }
\label{CE_3}
\end{figure}

\begin{figure}[h] 
	\centering
	\includegraphics[width=3.5in]{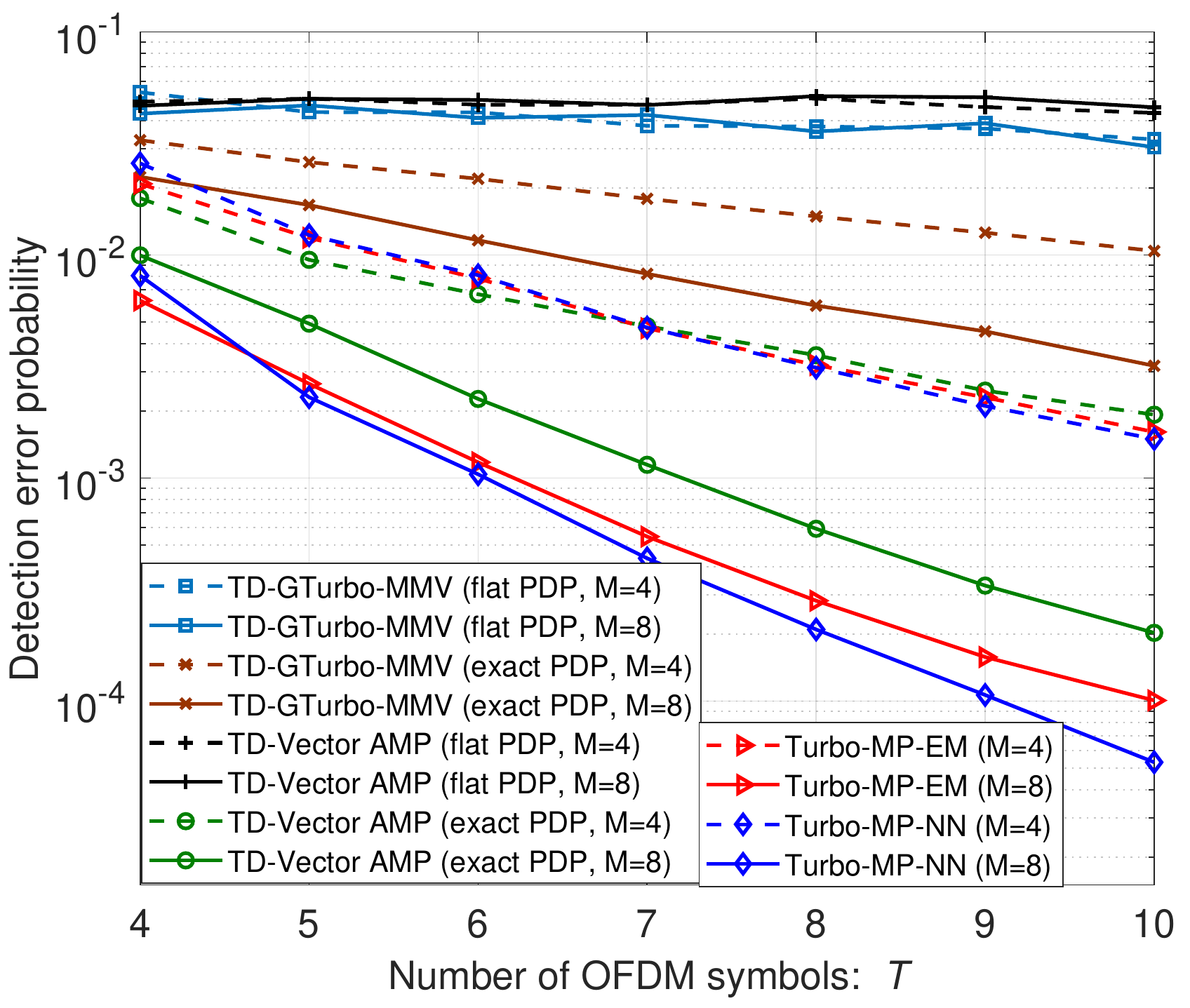}
	\caption{Detection error probability versus the number of OFDM symbols $T$. ${\rm SNR} = -15 $ dB and the sub-block number $Q=4$.}
\label{CE_5}
\end{figure}

Fig. \ref{CE_3} shows the channel estimation NMSE against the different number of OFDM symbols. There is a clear performance gap between Turbo-MP and the baselines at different $T$. Moreover, to reach $\rm{NMSE}=-15$ dB, the pilot overhead ($T=5$) for Turbo-MP is only half of that  ($T=10$) for TD-Vector AMP and TD-Gturbo-MMV with exact PDP. It is shown that our proposed scheme can support mMTC with a dramatically reduced overhead. In Fig. \ref{CE_5}, the detection error probability versus OFDM symbols is shown. The trend is similar to Fig. \ref{CE_4}. Turbo-MP achieves the best ADD performance at the different number of OFDM symbols.
% In conclusion, 

\begin{figure}[h] 
	\centering
	\includegraphics[width=3.5in]{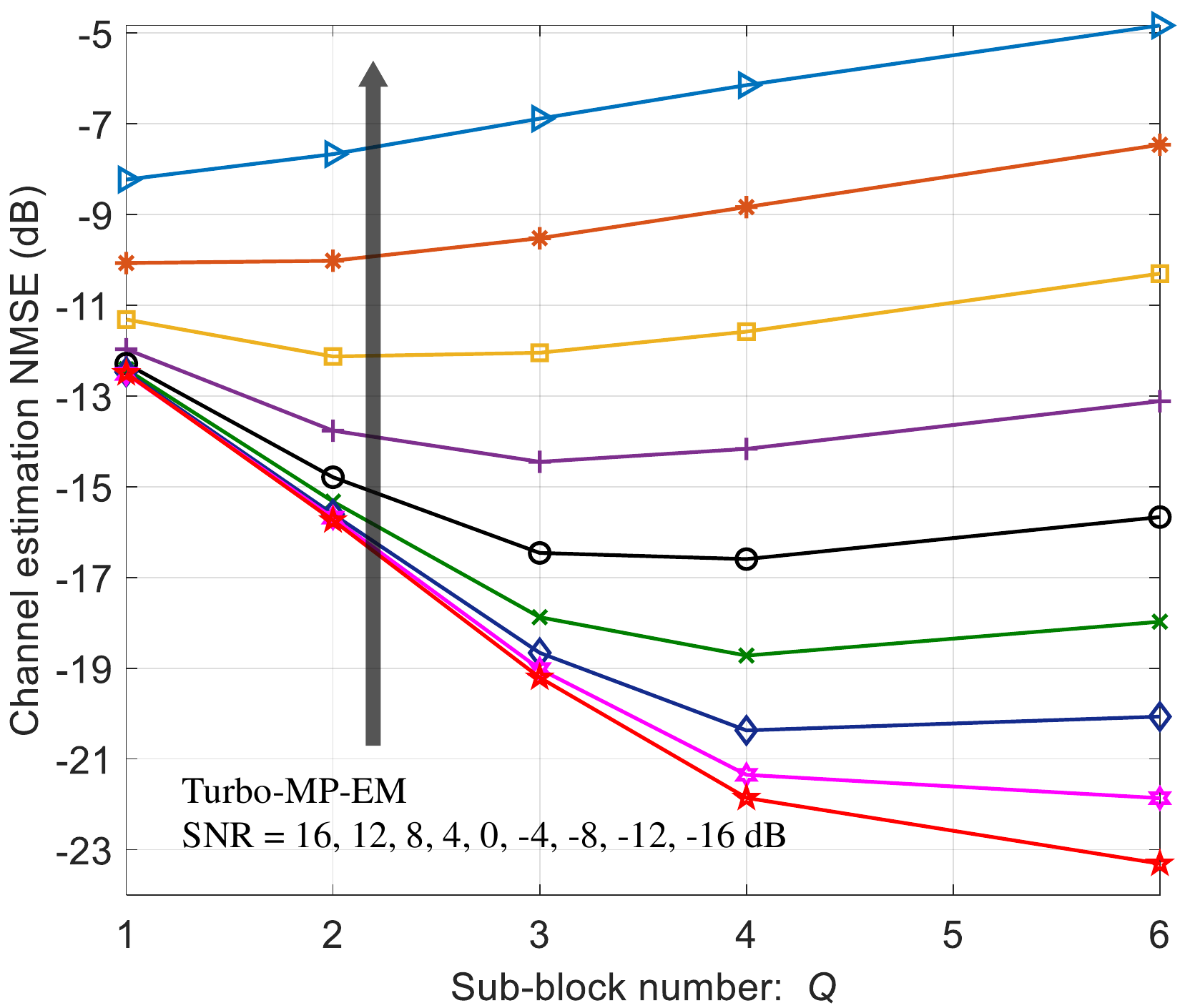}
	\caption{Channel estimation NMSE versus sub-block number $Q$ at different ${\rm SNR}$. The number of BS antennas $M=8$ and the number of OFDM symbols $T=10$. }
\label{CE_6}
\end{figure}

\begin{figure}[h] 
	\centering
	\includegraphics[width=3.5in]{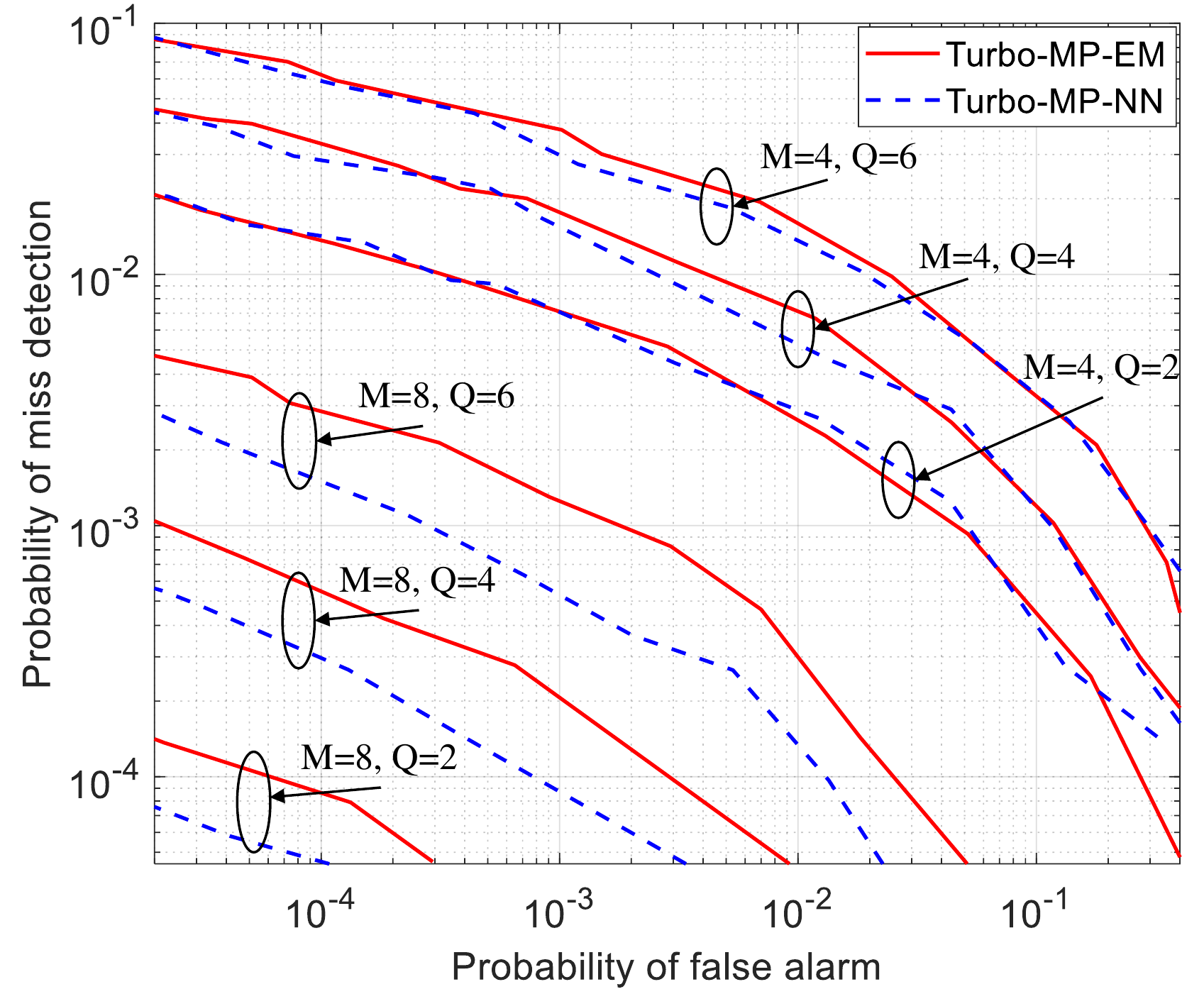}
	\caption{Probability of miss detection versus probability of false alarm at different sub-block number and antennas' number. ${\rm SNR} = -15 $ dB and the number of OFDM symbols $T=10$.}
\label{CE_7}
\end{figure}
Fig. \ref{CE_6} illustrates the impact of the sub-block number on the CE performance at different SNR, which implies the trade-off between the estimation accuracy and the model accuracy. In specific, it is seen that at low SNR, a smaller sub-block number corresponds to better NMSE performance while the case is contrary at high SNR. The reason is that the NMSE performance is mainly affected by AWGN at low SNR, in which case a small sub-block number helps to improve the estimation accuracy. However, at high SNR, the CE performance is limited by the model mismatch which can be reduced by increasing the sub-block number. In practice, a proper sub-block number needs to be chosen according to the wireless environment. Fig. \ref{CE_7} shows the miss detection probability versus false alarm probability, where we modify the threshold $\lambda^{thr}$ to reach the trade-off between miss detection and false alarm. Clearly, Turbo-MP-NN outperforms Turbo-MP-EM at different settings of sub-block number and antennas number. Besides, we see that Turbo-MP with the smaller sub-block number corresponds to a better ADD performance at ${\rm SNR}=-15$ dB. This result is consistent with the CE performance in Fig. \ref{CE_6}. From Fig. \ref{CE_6} and Fig.\ref{CE_7}, we find that the algorithms can achieve excellent device detection performance at relative low SNR while the accurate channel estimation requires a higher SNR. Such observation suggests that when the BS is equipped with multiple antennas, the bottleneck in mMTC is CE instead of ADD.

\section{Conclusion}
In this paper, a frequency-domain block-wise linear channel model was established in 
the MIMO-OFDM-based grant-free NOMA system to effectively compensate the 
channel frequency-selectivity and reduce the number with a small number of variables 
to be determined in channel estimation. From the perspective of Bayesian inference, 
we designed the low-complexity Turbo-MP algorithm to solve the ADD and CE 
problem, where machine learning was incorporated to learn the prior parameters. We 
numerically show that Turbo-MP designed for the proposed block-wise linear model 
significantly outperforms the state-of-the-art counterpart algorithms.

\ifCLASSOPTIONcaptionsoff
  \newpage
\fi



%\bibliographystyle{IEEEtran}
%\begin{thebibliography}{1}
%\end{thebibliography}

\bibliographystyle{IEEEtran}
\bibliography{TurboMP}

%\end{IEEEbiography}
%% insert where needed to balance the two columns on the last page with
%%\newpage


%\vfill


\end{document}